\documentclass[%
 reprint,
 amsmath,amssymb,
 aps,
]{revtex4-2}

\usepackage{graphicx}
\usepackage{dcolumn}
\usepackage{bm}
\usepackage{color,soul}
\usepackage{amsmath}
\usepackage[utf8]{inputenc}
\usepackage[english]{babel}
\usepackage{setspace}

\usepackage{chngcntr}
\counterwithout{figure}{section}
\counterwithout{equation}{section}

\begin{document}


\title{Advanced Phase-Change Materials for Enhanced Meta-Displays}

\author{Omid Hemmatyar$^{1}$}
\author{Sajjad Abdollahramezani$^{1}$}
\author{Sergey Lepeshov$^{2}$}
\author{Alex Krasnok$^{3}$}
\author{Tyler Brown$^{1}$}
\author{Andrea Al{\`u}$^{3, 4}$}
\author{Ali Adibi$^{1}$}

\affiliation{$^{1}$School of Electrical and Computer Engineering, Georgia Institute of Technology, 778 Atlantic Drive NW, Atlanta, Georgia 30332-0250, United States}
\affiliation{$^{2}$ITMO University, St. Petersburg 197101, Russia}
\affiliation{$^{3}$Photonics Initiative, Advanced Science Research Center, City University of New York, New York, NY 10031, United States}
\affiliation{$^{4}$Physics Program, Graduate Center, City University of New York, New York, NY 10016, United States}

\date{\today}

\begin{abstract}

Structural colors generated due to light scattering from static all-dielectric metasurfaces have successfully enabled high-resolution, high-saturation and wide-gamut color printing applications. Despite recent advances, most demonstrations of these structure-dependent colors lack post-fabrication tunability that hinders their applicability for front-end dynamic display technologies. Phase-change materials (PCMs), with significant contrast of their optical properties between their amorphous and crystalline states, have demonstrated promising potentials in reconfigurable nanophotonics. Herein, we leverage a tunable all-dielectric reflective metasurface made of a newly emerged class of low-loss optical PCMs with superb characteristics, i.e., antimony trisulphide (Sb$_2$S$_3$), antimony triselenide (Sb$_2$Se$_3$), and binary germanium-doped selenide (GeSe$_3$), to realize switchable, high-saturation, high-efficiency and high-resolution structural colors. Having polarization sensitive building blocks, the presented metasurface can generate two different colors when illuminated by two orthogonally polarized incident beams. Such degrees of freedom (i.e., structural state and polarization) enable a single reconfigurable metasurface with fixed geometrical parameters to generate four distinct wide-gamut colors suitable for a wide range of applications, including tunable full-color printing and displays, information encryption, and anti-counterfeiting.

\end{abstract}

\keywords{dynamic metasurfaces, phase-change materials, nanophotonics, structural colors}

\maketitle


\section*{Introduction}

In the past decades, absorption and emission of light from organic dyes and chemical pigments have been the most common color generation mechanisms in color-imaging and display devices \cite{daqiqeh2020nanophotonic}. Nevertheless, there are still several challenges with the developed technologies, such as environmental hazards, vulnerability to high-intensity light, and limited scalability to smaller pixel sizes. In order to address these issues, structural colors emerged as compelling alternatives. Structural colors are observed in numerous natural species whose colors arise from light scattering and interference in micro/nanostructured patterns of their skins or scales \cite{vukusic1999quantified}. Inspired by nature and considering the recent advancement in nanofabrication, artificial structural colors generated via a resonant interaction between the incident white light and building blocks in optical metasurfaces \cite{alu2005achieving, krasnok2012all, yu2014flat, miroshnichenko2010fano, decker2015high, abdollahramezani2015analog,  kuznetsov2016optically, hemmatyar2017phase, hu2020moire, arik2020beam, hemmatyar2020wide, abdollahramezani2020meta}, i.e., arrays of subwavelength patterned nanostructures, have gained great attention in recent years. In this context, plasmonic metasurfaces made of gold, silver and aluminum nanostructures have been extensively used to generate structural colors based on plasmon resonances \cite{kumar2012printing, tan2014plasmonic, kristensen2016plasmonic, rezaei2019wide}. Despite their versatility, the broad and weak plasmon resonances, imposed by the significant inherent ohmic loss of the constituent metallic materials, result in low color saturation and purity \cite{kristensen2016plasmonic}.

\begin{figure*}[htbp]
\centering
\includegraphics[width=1\linewidth, trim={0cm 0cm 0cm 0cm},clip]{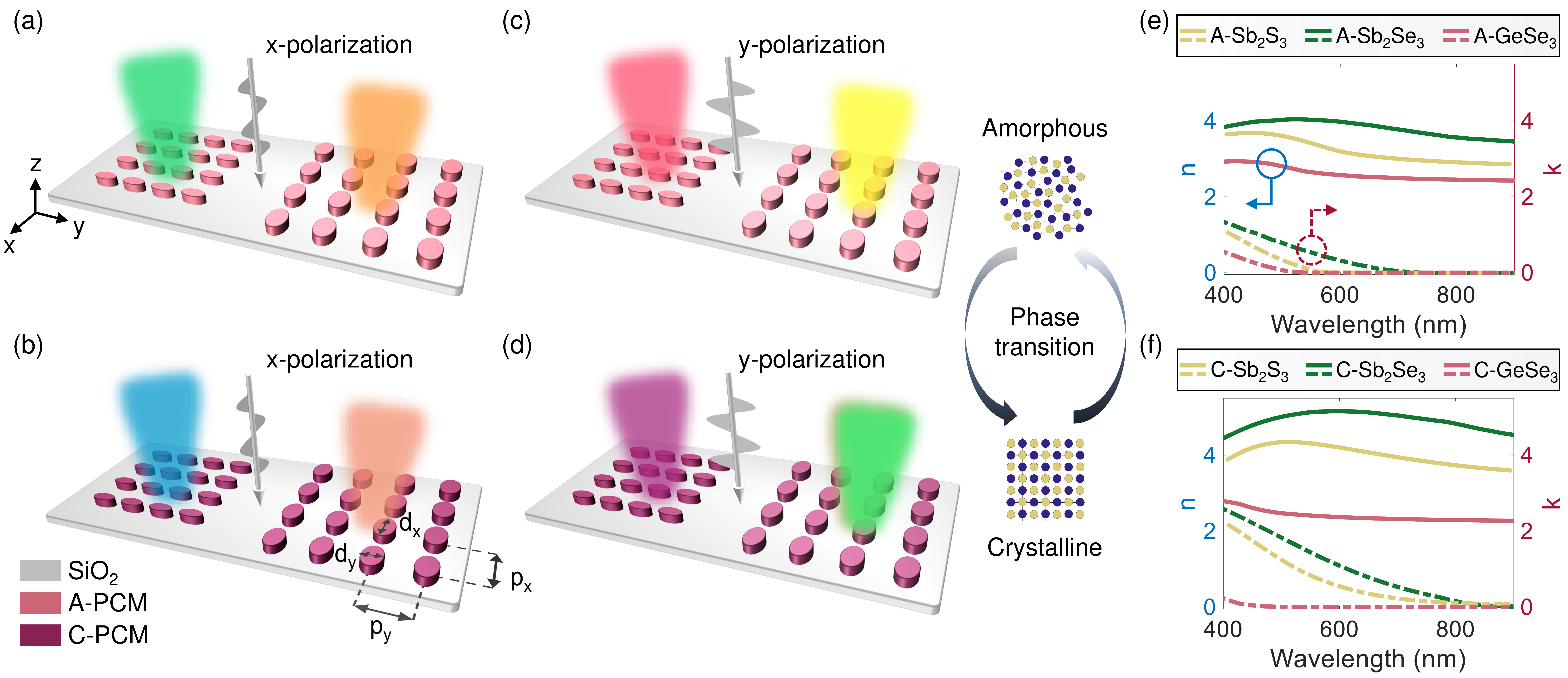}
\caption{\textbf{Multicolor generation by polarization-sensitive tunable reflective metasurfaces consisting of elliptical nanopillars made of low-loss PCMs on a glass substrate.} Each metasurface can generate four different colors, two of them are attributed to the using of \textbf{(a,b)} x-polarized and \textbf{(c,d)} y-polarized incident white light, and two others correspond to the (a,c) amorphous and (b,d) crystalline phases of the constituent nanopillars. For all metasurfaces, the height ($h$) of the nanopillars are fixed while the periodicities in x and y directions (i.e., $p_x$ and $p_y$, respectively) change to generate different colors. The major and minor axes of the nanopillars in x and y directions are proportional to the periodicity in that direction with a constant aspect ratio, i.e. $d_{x,y} = \alpha \, p_{x,y}$, in which $\alpha$ is constant. \textbf{(e,f)} Real and imaginary parts of the refractive index of Sb$_2$S$_3$, Sb$_2$Se$_3$, and GeSe$_3$ for (e) amorphous (A) and (f) crystalline (C) phases. The colors shown in (a-d) correspond to two GeSe$_3$ metasurfaces with $p_x=290$ nm and $p_y=410$ nm (in the left metasurface), and $p_x=390$ nm and $p_y=350$ nm (in the right metasurface), both with $h=250$ nm and $\alpha = 0.55$.}
\label{Fig_1}
\end{figure*}

To meet the challenges associated with plasmonic metasurfaces, recently, all-dielectric metasurfaces made of high-refractive-index materials supporting electric dipole (ED) and magnetic dipole (MD) Mie-type resonances have been used for generating a full range of vivid and highly saturated structural colors desired for high-resolution display technologies \cite{shen2015structural, sun2017all, zhu2017resonant, dong2017printing, jin2018noninterleaved, yang2019ultrahighly, hemmatyar2019full, yang2020all, daqiqeh2020nanophotonic, hemmatyar2020fano}. However, these colors are fixed-by-design and cannot be tuned since the geometrical parameters of passive all-dielectric metasurfaces remain unchanged after fabrication. In order to enable active display applications, a real-time color tunability is essential.

To realize high-resolution structural color tunability in metasurfaces, several modulation techniques have been proposed. Using liquid crystals in conjunction with plasmonic nanoantennae \cite{franklin2015polarization, olson2016high}, utilizing mechanically stretchable substrates integrated with plasmonic \cite{tseng2017two, song2017actively} and dielectric \cite{gutruf2016mechanically, quan2020stretchable} nanoscatterers, changing the refractive index of the medium surrounded nanostructures \cite{king2015fano}, modifying the optical properties of the constituent magnesium-based nano-cavities of a hybrid plasmonic-dielectric platform via a chemical reaction \cite{chen2017dynamic}, and changing the polarization state of incident light \cite{yang2018polarization}.
Despite impressive advancements, these approaches can hardly meet the requirements for lightweight, flexible, durable, and cost-effective dynamic reflective color displays with high color contrast and saturation, multiple stable colors, and high refreshing rates. 

To overcome these shortcomings, chalcogenide phase-change materials (PCMs) \cite{wuttig2017phase,feldmann2019all,  ding2019dynamic, abdollahramezani2020tunable, hail2019optical, dong2018tunable, gholipour2013all, michel2013using, abdollahramezani2018reconfigurable, zhang2019broadband, taghinejad2020ito, wang2016optically, rios2015integrated, abdollahramezani2021electrically, krasnok2020active, abdollahramezani2018dynamic, zheng2018gst, tian2019active, michel2019advanced, xu2019low, abdollahramezani2021dynamic,wu2021programmable, hemmatyar2020mixed, de2018nonvolatile, feldmann2021parallel, abdollahramezani2020electrically, zheng2020nonvolatile, leitis2020all, hemmatyar2020tunable2, zhu2020linear, zhu2021tunable, shalaginov2021reconfigurable, zhang2021electrically, wang2021electrical, lepeshov2021tunable}, whose optical properties (e.g., refractive index) can be remarkably modified upon applying an external stimulus (optically, electrically or thermally), have been successfully used as the tunable materials for color switching \cite{hosseini2014optoelectronic, schlich2015color, tao2020phase, yoo2016multicolor, jafari2019reconfigurable, carrillo2019nonvolatile, de2020reconfigurable, hemmatyar2020tunable}. The advantages of PCM-based color-switching techniques over other counterparts originate from unique electrical and optical features of PCMs including ultrafast reversible switching speed (10s -100s nanoseconds) between two stable phases with high durability (up to $10^{15}$ cycles), notable scalability (down to nm-scale sizes), good thermal stability (up to several hundred degrees), and adaptability with CMOS fabrication technology. Considering these unique features, single ultrathin film \cite{hosseini2014optoelectronic, schlich2015color, tao2020phase} or multiple ultrathin films \cite{yoo2016multicolor, jafari2019reconfigurable} made of germanium antimony telluride (GeSbTe or GST in short) and germanium telluride (GeTe) alloys in a multistack configuration with other dielectric and/or metallic films have been utilized for color switching \cite{carrillo2019nonvolatile}. In spite of unparalleled properties of PCMs, these approaches suffer from the high absorption loss of GST and GeTe within the visible wavelength range, which results in low-quality-factor (low-Q) reflectance resonances. This, in turn, also yields colors with low saturation and low color value (i.e., the reflectance value at the resonance peak) and purity in both amorphous and crystalline states of these PCMs.

To address these challenges and enable next-generation color displays, here, we report the design strategy and implementation of a tunable all-dielectric metasurface consisting of elliptical PCM nanopillars formed in antimony trisulphide (Sb$_2$S$_3$), antimony triselenide (Sb$_2$Se$_3$), or binary germanium-doped selenide (GeSe$_3$) all of which having low absorption loss and wide bandgap within the visible window \cite{dong2019wide, chen2015optical, ghazi2019strong, delaney2020new, delaney2021non}. Since the real part of the refractive index of these materials in their amorphous and crystalline states are large enough to support Mie-type ED and MD resonances with refractive-index-sensitive spectral positions, our technique can enable high-resolution phase-transition-based color switching with high saturation and purity. Moreover, owing to the polarization-sensitivity of the asymmetric elliptical PCM nanopillars of the presented metasurfaces, we can encode two different colors into two orthogonal polarization states of the incident light. Therefore, in our technique, one metasurface with fixed geometrical parameters can generate four different colors upon transition in the structural state of the contributed PCM. This ability enables many interesting applications, including tunable full-color printing and displays, information encryption, anticounterfeiting, wearable screens, and electronic paper.

\section*{Results and Discussion}

Figure~\ref{Fig_1}a demonstrates two all-dielectric reflective metasurfaces, each consisting of periodic arrays of asymmetric elliptical PCM nanopillars on top of a glass substrate with rectangular unit cells with different periodicities along x- and y-directions  (i.e., $p_x$ and $p_y$ in Figure~\ref{Fig_1}b). The major and minor axes of the PCM nanopillars are proportional to the periodicities of the unit cell in the corresponding direction, i.e., $d_{x,y} = \alpha \, p_{x,y}$, in which $\alpha$ is a constant between 0 and 1. The height of the nanopillars ($h$) is constant for fabrication preference. The angle between the linear polarization of the normally incident light and the x-axis is denoted as $\varphi$. The normally incident x-polarized white light ($\varphi = 0^\circ$) can be reflected in various colors by varying the geometrical parameters of the elliptic amorphous-PCM (A-PCM) nanopillars and those of the unit cell (see Figure~\ref{Fig_1}a). Upon phase transition, crystalline-PCM (C-PCM) nanopillars with the same geometrical parameters and under the same illumination conditions generate colors that are different from those in the A-PCM (see Figure~\ref{Fig_1}b). This phase-transition-based color switching is attributed to the refractive index change of the constituent PCMs upon their transition between amorphous and crystalline phases. The PCM in these structures can be Sb$_2$S$_3$, Sb$_2$Se$_3$, or GeSe$_3$ due to their wide-bandgap and low optical loss within the visible wavelength range. The refractive index ($n$) and the extinction coefficient ($k$) of these materials in the amorphous and crystalline phase-states are shown in Figures~\ref{Fig_1}e,f, respectively.

In addition to the phase-transition-based color-switching mechanism mentioned above, the circularly asymmetric PCM nanopillars can enable a polarization-based color-switching mechanism in which same metasurfaces can generate different colors when illuminated by x- and y-polarized ($\varphi = 90^\circ$) incident white light for both A-PCM (Figures~\ref{Fig_1}a,c) and C-PCM (Figures~\ref{Fig_1}b,d) nanopillars. Therefore, one metasurface with fixed geometrical parameters can generate four different colors owing to the phase-state-tunability (two colors) and polarization-sensitivity (two colors) of the constituent PCM nanopillars.

\begin{figure*}[htbp]
\centering
\includegraphics[width=1\linewidth, trim={0cm 0cm 0cm 0cm},clip]{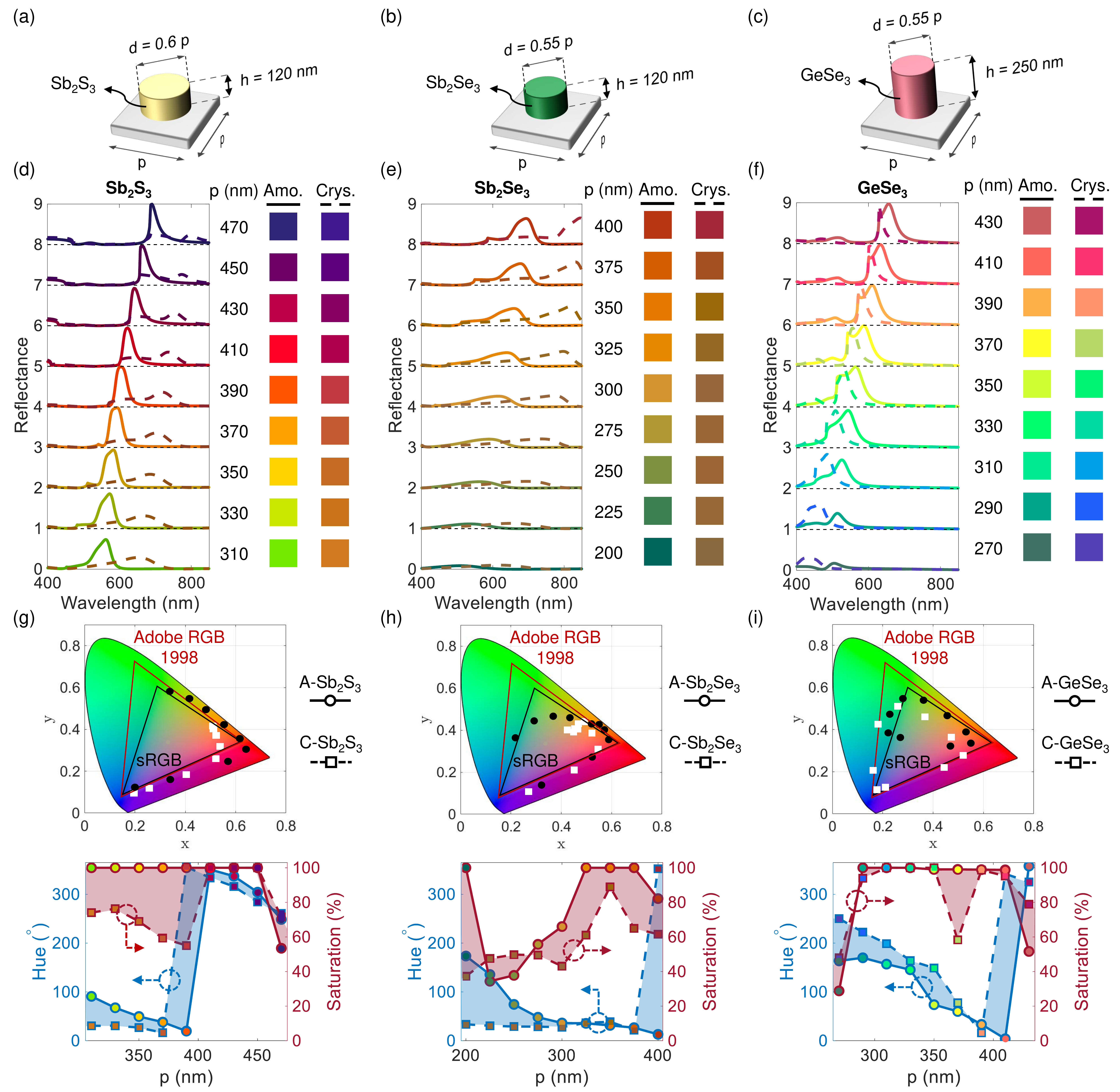}
\caption{\textbf{Color switching enabled by phase-transition of the PCM nanopillars.} \textbf{(a-c)} Schematic and geometrical parameters of a unit cell of a polarization-insensitive PCM metasurface made of (a) Sb$_2$S$_3$, (b) Sb$_2$Se$_3$ and (c) GeSe$_3$ circular nanopillars with a fixed heigh $h$. The periodicity of the unit cell in both x and y directions is $p$, and the diameter of the nanopillars is $d = \alpha \, p$ with $\alpha$ being a constant. \textbf{(d-f)} Simulated reflectance spectra for the amorphous (solid lines) and crystalline (dashed lines) phases and their corresponding colors for different periodicities ($p$). The PCM is (d), (e), and (f) is Sb$_2$S$_3$, Sb$_2$Se$_3$, and GeSe$_3$, respectively. The curves for different $p$s are diplaced vertically for better visibility and comparison. The sharp resonances observed in (d-f) are attributed to the interference between ED and MD modes inside the PCM nanopillars. Upon the PCM phase transition, a red-shift of $|\Delta\lambda_{\textrm{Sb}_{2}\textrm{S}_{3}}|>180$ nm and $|\Delta\lambda_{\textrm{Sb}_{2}\textrm{Se}_{3}}|>200$ nm is observed for the case of (d) Sb$_2$S$_3$ and (e) Sb$_2$Se$_3$, respectively, while a blue-shift of $|\Delta\lambda_{\textrm{Ge}\textrm{Se}_{3}}|<70$ nm is observed for the case of (f) GeSe$_3$. \textbf{(g-i)} Corresponding CIE 1931 chromaticity coordinates of the reflectance spectra, and the hue and saturation values of the colors shown in (d-f) for amorphous (black circles in the top panel and circle-solid line in bottom panel) and crystalline (white squares in the top panel and square-dashed line in the bottom panel) phases of the corresponding PCMs in (d-f).}
\label{Fig_2}
\end{figure*}

To study only the effect of phase transition of the constituent PCM nanopillars on the reflected color, we first consider metasurfaces with a square lattice ($p_x=p_y=p$) of circular ($d_x=d_y=d$) PCM nanopillars for three cases of Sb$_2$S$_3$, Sb$_2$Se$_3$, or GeSe$_3$ as shown in Figures~\ref{Fig_2}a-c, respectively. To cover a full range of colors, we used the geometrical parameters in Figures~\ref{Fig_2}a-c and change the periodicity of the unit cell from $p=310$ nm to $p=470$ nm with a step of 20 nm for Sb$_2$S$_3$, from $p=200$ nm to $p=400$ nm with a step of 25 nm for Sb$_2$Se$_3$, and from $p=270$ nm to $p=430$ nm with a step of 20 nm for GeSe$_3$, and obtain the corresponding reflectance spectra, and in turn, the generated colors as shown in Figures~\ref{Fig_2}d-f, respectively. The method used for obtaining the colors from their corresponding reflectance spectra is explained in Supporting Information Note~I. The full-wave simulations of the reflectance response of the metasurfaces are performed using the commercial software Lumerical Solutions based on the finite-different time-domain (FDTD) technique. The periodic boundary condition is used in x- and y-directions to mimic the periodicity, while perfectly matched layers are used in the z-direction (top and bottom layers) to model the free space. The refractive index of the glass substrate is 1.46.

As shown in Figures~\ref{Fig_2}d-f, by increasing $p$, the spectral position of the reflectance resonances for both amorphous (solid lines) and crystalline (dashed lines) states red-shifts. The spectral position of each of these resonances is dependent on the refractive index of the constituent PCM owing to the interference between ED and MD modes inside the PCM nanopillars as will be discussed later. Therefore, by switching the state of the nanopillars from amorphous to crystalline, the central wavelengths of the resonances red-shift in the cases of Sb$_2$S$_3$ and Sb$_2$Se$_3$ (see Figures~\ref{Fig_2}d,e), and blue-shift in the case of GeSe$_3$ (see Figure~\ref{Fig_2}f) because $\Delta n_{\textrm{Sb}_{2}\textrm{S}_{3}}$, $\Delta n_{\textrm{Sb}_{2}\textrm{Se}_{3}} > 0$, while $\Delta n_{\textrm{GeSe}_{3}} < 0$ (with $\Delta n = n_{\textrm{C-PCM}} - n_{\textrm{A-PCM}}$) within the visible wavelength range (see the refractive indices in Figures~\ref{Fig_1}e,f). The actual shift of the resonance wavelengths are $|\Delta\lambda_{\textrm{Sb}_{2}\textrm{S}_{3}}| < 180$ nm, $|\Delta\lambda_{\textrm{Sb}_{2}\textrm{Se}_{3}}| < 200$ nm, and $|\Delta\lambda_{\textrm{GeSe}_{3}}| < 70$ nm (see Figures~\ref{Fig_2}d-f). The relative strength of the wavelength shifts in these PCMs, i.e. $|\Delta\lambda_{\textrm{Sb}_{2}\textrm{Se}_{3}}| > |\Delta\lambda_{\textrm{Sb}_{2}\textrm{S}_{3}}| > |\Delta\lambda_{\textrm{GeSe}_{3}}|$, is attributed to the relative strength of the change in the real part of their refractive indices upon the phase transition between amorphous and crystalline, i.e. $|\Delta n_{\textrm{Sb}_{2}\textrm{Se}_{3}}| > |\Delta n_{\textrm{Sb}_{2}\textrm{S}_{3}}| > |\Delta n_{\textrm{GeSe}_{3}}|$, as shown in Figure~S1 in the Supporting Information. On the other hand, the sharpness of the reflectance resonances in Figures~\ref{Fig_2}d-f is mainly dependent on the PCM extinction coefficient shown as the dashed curves in Figures~\ref{Fig_1}e,f. In the case of Sb$_2$S$_3$ nanopillars, the high-efficiency resonances (i.e., those with high reflectance value at the resonance peak) in the low-loss amorphous phase are damped upon the transition to the crystalline phase with higher absorption loss (compare solid and dashed curves in Figure~\ref{Fig_2}d). This high absorption loss arises for both amorphous and crystalline Sb$_2$Se$_3$ nanopillars, resulting in relatively low-efficiency reflectance resonances (see Figure~\ref{Fig_2}f). In contrast, GeSe$_3$ nanopillars remain very low-loss across the entire visible range for both the amorphous and crystalline phases, yielding high-efficiency resonances in both cases (see Figure~\ref{Fig_2}f).

For a direct comparison between the presented three PCMs in terms of color generation/switching, Figures~\ref{Fig_2}g-i show the generated colors in the amorphous (black circles) and crystalline (white squares) phases in the same International Commission on Illumination (CIE) 1931 chromaticity coordinates for the three PCMs in top panels, and their corresponding hue and saturation values for amorphous (solid-circle lines) and crystalline (dashed-square curves) phases in the bottom panels. The approach of calculating the CIE XYZ tristimulus of the reflectance spectra and their corresponding hue and saturation values are given in the Supporting Information Note~I. In terms of the color gamut coverage, the calculated color gamut area for A-Sb$_2$Se$_3$ (C-Sb$_2$Se$_3$) is around 98.3\% (43.4\%) of the standard RGB (sRGB) and 72.9\% (32.2\%) of the Adobe RGB, from Figure~\ref{Fig_2}g. The color gamut area for the case of A-Sb$_2$Se$_3$ (C-Sb$_2$Se$_3$) is around 70.1\% (33.3\%) of the sRGB, and 52\% (24.7\%) of the Adobe RGB (from Figure~\ref{Fig_2}h). For the case of A-GeSe$_3$ (C-GeSe$_3$), a full-range of colors with gamut area of 57.8\% (90.8\%) of the sRGB, and 42.9\% (67.3\%) of the Adobe RGB can be obtained (from Figure~\ref{Fig_2}i). Therefore, in terms of color gamut area, Sb$_2$S$_3$ and GeSe$_3$ have almost the same performance, yet better than Sb$_2$Se$_3$. Moreover, these results show that our all-dielectric PCM-based metasurfaces can generate a wide color gamut larger than the state-of-the-art plasmonic colors ($\sim45\%$ of sRGB \cite{rezaei2019wide}) for A-Sb$_2$S$_3$, A-Sb$_2$Se$_3$ and A/C-GeSe$_3$ cases.

\begin{figure*}[htbp]
\centering
\includegraphics[width=.82\linewidth, trim={0cm 0cm 0cm 0cm},clip]{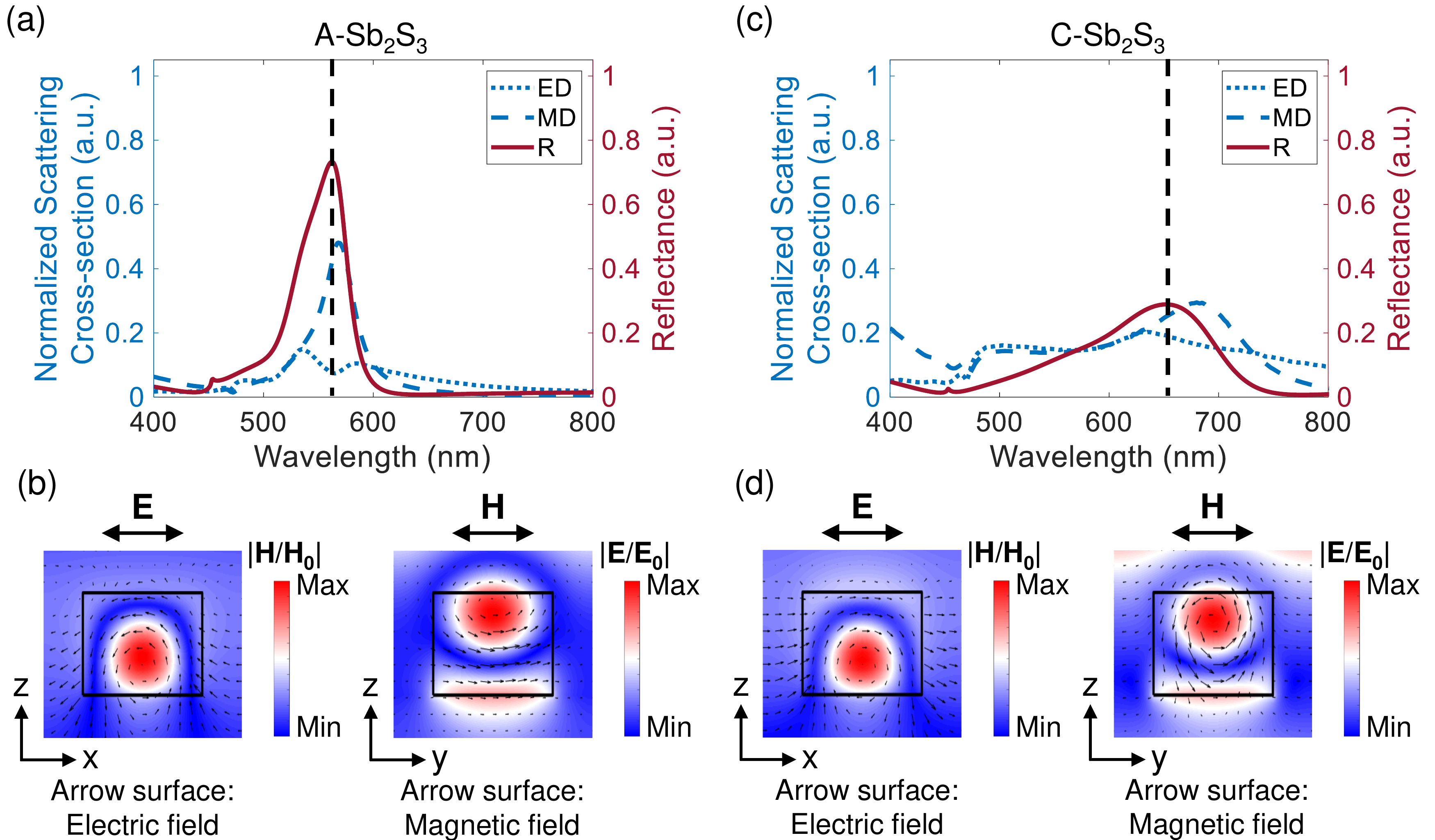}
\caption{\textbf{Multipolar decomposition analysis.} \textbf{(a,c)} Calculated normalized scattering cross-sections and simulated reflectance (R) spectrum of a Sb$_2$S$_3$ metasurface with geometrical parameters of $p=310$ nm, $d=0.6 \, p$, and $h=120$ nm for the (a) amorphous and (c) crystalline phases. The constructive interference between the ED and MD modes at $\lambda_a = 560$ nm ($\lambda_c = 652$ nm) boosts the backward scattering intensity, and in turn, results in a reflectance peak in the case of A-Sb$_2$S$_3$ (C-Sb$_2$S$_3$). \textbf{(b,d)} Normalized magnetic field intensity with arrow surface of electric field (left panel), and normalized electric field intensity with arrow surface of magnetic field (right panel) for the metasurfaces in (a) and (c) at (b) $\lambda_a = 560$ nm and (d) $\lambda_c = 652$ nm, respectively.} 
\label{Fig_22}
\end{figure*}

In the RGB color-mixing model, the hue (H) is defined as the proportion of the dominant wavelength (resonance wavelength in this case) with respect to other wavelengths in the reflected light and is independent of the intensity of the light. It simply indicates the "perceived color" by the human eyes and ranges from $0^\circ$ to $360^\circ$, in which $0^\circ$ (and $360^\circ$), $120^\circ$ and $240^\circ$ represent pure red, pure green, and pure blue, respectively (See Figure~S2 in the Supporting Information for more details). The saturation, on the other hand, is defined as the ratio of the intensity of the reflected light at the resonance wavelength (associated to the perceived color) to that of the incident white light, simply indicating the purity of a color and ranging from 0\% to 100\%. Considering this definition, the narrower the bandwidth of the reflectance resonance, the higher the saturation of the generated color. In the content of color switching between two phases, the performance measure is achieving two highly-saturated colors in both phases with a maximum hue variation ($\Delta \textrm{H} = \textrm{H}_{\textrm{C-PCM}} - \textrm{H}_{\textrm{A-PCM}}$) upon switching. To analyze the performance of the presented phase-transition-based color-switching approach, the hue and saturation values of the simulated colors in Figures~\ref{Fig_2}d-f are plotted in the bottom panels in Figures~\ref{Fig_2}g-i for both amorphous (solid-circle lines) and crystalline (dashed-square curves) phases of the PCMs. In terms of saturation preservation upon phase transition, GeSe$_3$ shows high-saturation values for both amorphous and crystalline cases (due to sharp reflectance resonances), while Sb$_2$S$_3$ shows highly saturated colors only in the amorphous phase. Sb$_2$Se$_3$, however, demonstrates a median level of saturation values in both amorphous and crystalline phases due to the wide reflectance resonances. With regards to hue variation, the hues of the generated colors in Sb$_2$S$_3$ and GeSe$_3$ cases change by varying $p$ in both amorphous and crystalline states while maintaining $\Delta \textrm{H} < 80^\circ$ upon phase transition. One may use this feature to switch the coloration of pixels of an image individually with each pixel being a Sb$_2$S$_3$ or GeSe$_3$ metasurface formed by of an array of down to 5$\times$5 or 6$\times$6 nanopillars \cite{hemmatyar2019full}. In the case of Sb$_2$Se$_3$, however, by changing $p$, all the varying hue values in the amorphous phase switch to an almost fixed hue in the crystalline phase. Using this property, one can turn off all the pixels of an image on a display comprising Sb$_2$Se$_3$ metasurfaces (i.e., pixels) by switching the phase of the Sb$_2$Se$_3$ nanopillars from the amorphous state (ON-state) to the crystalline state (OFF-state). This is a unique feature that is absent in other approaches in previous works, e.g., the polarization-sensitive color-switching approach \cite{yang2018polarization}.

\begin{figure*}[htbp]
\centering
\includegraphics[width=1\linewidth, trim={0cm 0cm 0cm 0cm},clip]{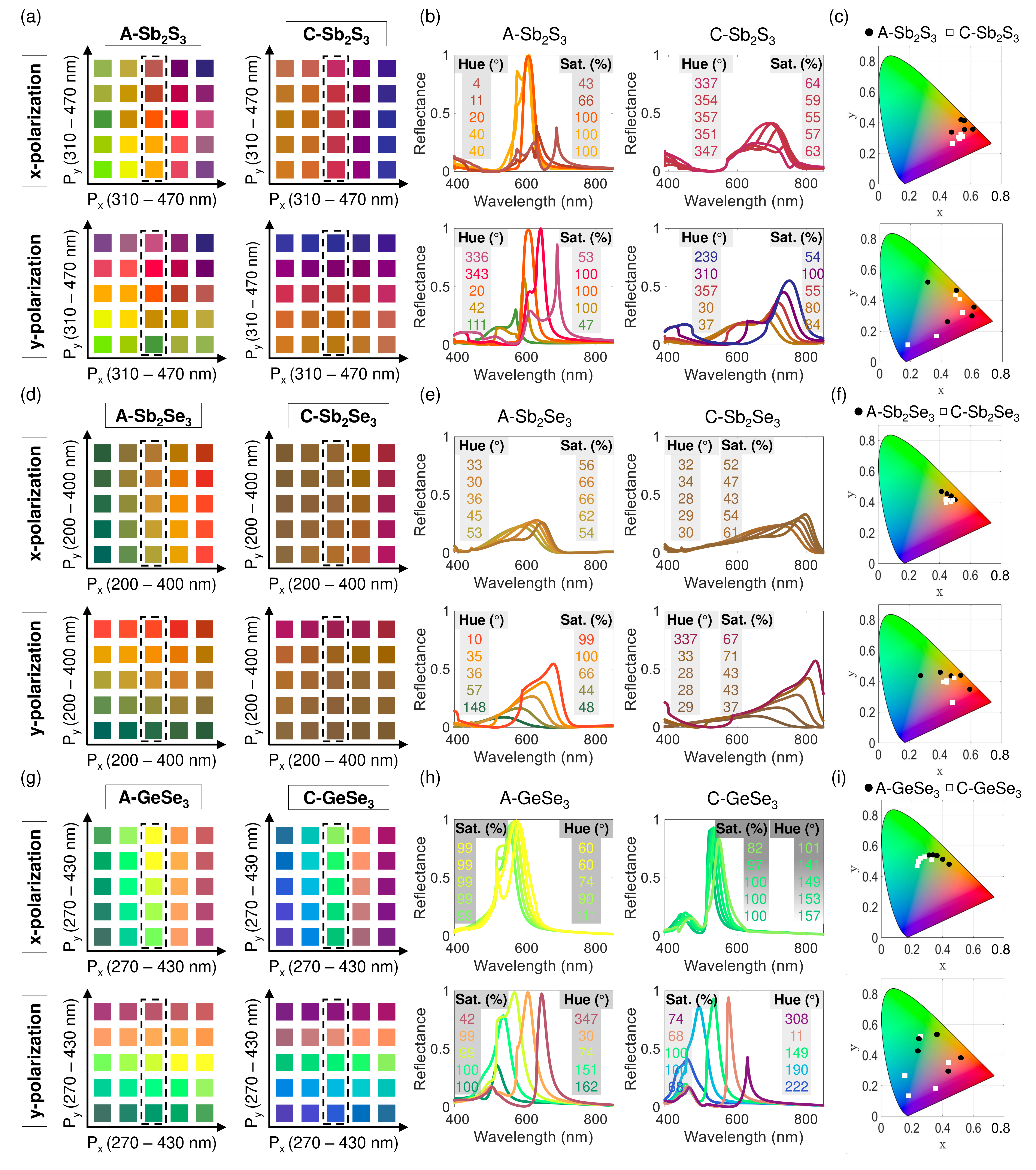}
\caption{\textbf{Multiple color generation enabled by phase-transition-based and polarization-based color switching mechanisms.} \textbf{(a,d,g)} Generated color palettes considering different periodicities in x- and y-directions ($p_x$ and $p_y$, respectively) for (a) Sb$_2$S$_3$ ($\alpha=0.6$ and $h=120$ nm), (d) Sb$_2$Se$_3$ ($\alpha=0.55$ and $h=120$ nm), and (g) GeSe$_3$ ($\alpha=0.55$ and $h=250$ nm). $p_x$ and $p_y$ in (a), (d) and (g) vary with 40 nm, 50 nm, and 40 nm increments, respectively. \textbf{(b,e,h)} Reflectance spectra of the colors indicated by the dashed rectangular boxes shown in the corresponding color palette in (a,d,g), respectively, with the values of hue and saturation (sat.) in the inset. \textbf{(c,f,i)} Corresponding color gamuts for amorphous (black circles) and crystalline (white squares) phases of the corresponding PCM in (a,d,g), respectively. In each figure, the upper (lower) panel represents the results related to x-polarization (y-polarization).}
\label{Fig_3}
\end{figure*}

\begin{figure*}[htb]
\centering
\includegraphics[width=.95\linewidth, trim={0cm 0cm 0cm 0cm},clip]{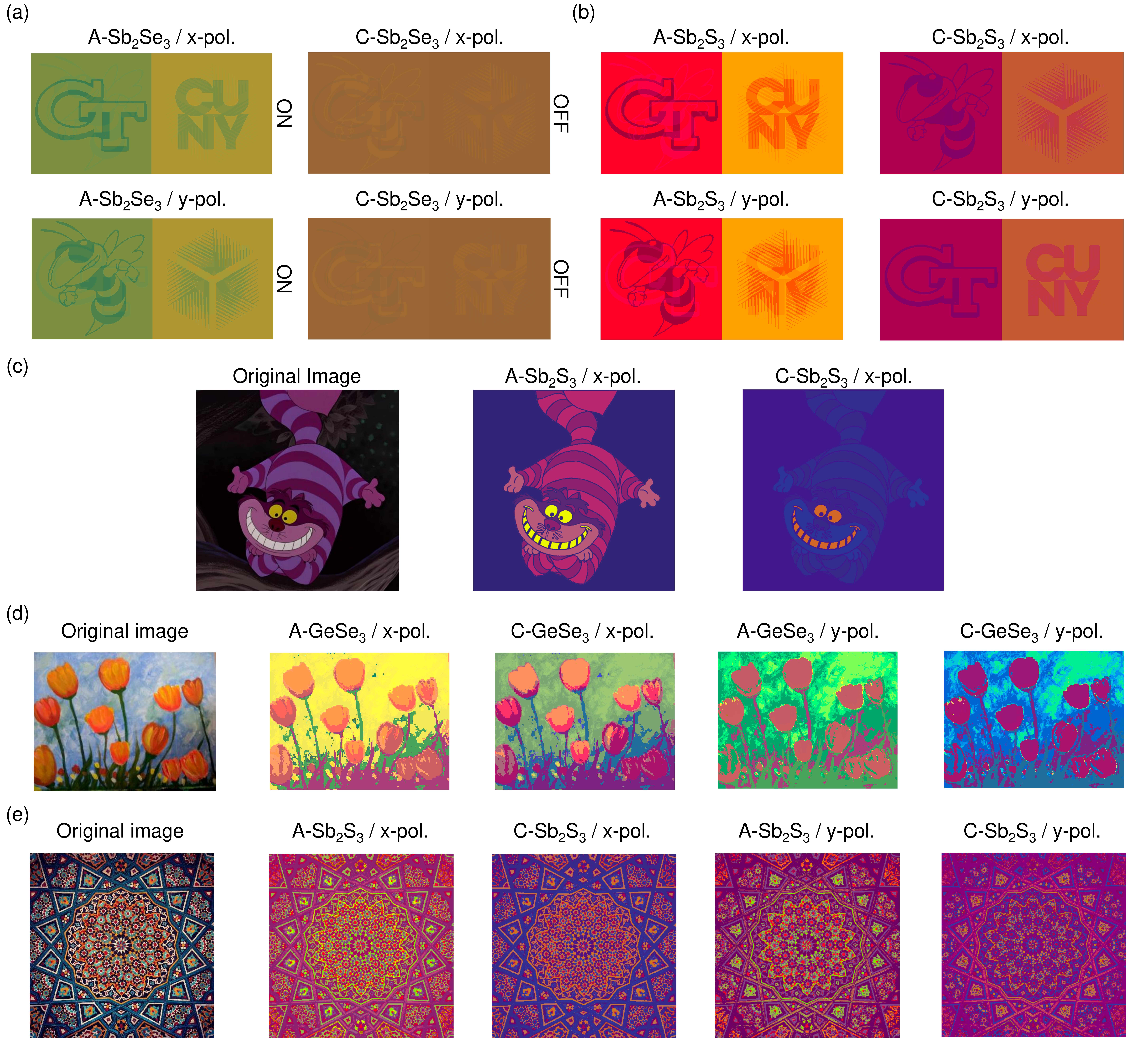}
\caption{\textbf{Multiple image encryption using PCM-based polarization-sensitive tunable metasurfaces.} \textbf{(a)} Encryption of four images (i.e., our institution logos and symbols) into an array of metasurfaces (color pixels) made of  Sb$_2$Se$_3$ nanopillars in the amorphous phase (ON-state) and crystalline phase (OFF-state) of Sb$_2$Se$_3$. The logos appear under x-polarization, while the symbols appear under y-polarization. Upon phase-transition into C-Sb$_2$Se$_3$, the images disappear (OFF-state). \textbf{(b)} Same images as those in (a) are reproduced by Sb$_2$S$_3$ metasurfaces in the amorphous phase. Upon phase-transition into C-Sb$_2$S$_3$, the logos (symbols) switched to the symbols (logos). \textbf{(c)} Reproducing of the image of Cheshire The Cat by A-Sb$_2$S$_3$ and C-Sb$_2$S$_3$ metasurfaces. Upon switching to the crystalline phase, all the body parts of Cheshire The Cat in the A-Sb$_2$S$_3$ case vanish except its teeth and eyes. \textbf{(d,e)} Reproduction of (d) a painting of flowers, and (e) an image of a Persian tile, using (d) GeSe$_3$ and (e) Sb$_2$S$_3$ metasurfaces. Switching the material phases and the incident polarization angles result in the regeneration of the original images in different colors resembling different time of the day in (d), and different patterns in (e).}
\label{Fig_5}
\end{figure*}

To investigate the physics behind the resonances that appeared in the reflectance spectra shown in Figures~\ref{Fig_2}d-f, we perform a multipole decomposition analysis on the electromagnetic response of the PCM nanopillars under white-light illumination as shown in Figure~\ref{Fig_22}. The description of multipole decomposition along with the result of this technique for other periods can be found in the Supporting Information Note~II and Figure~S3. Note that we apply periodic boundary conditions in x- and y-directions and hence, the multipole decomposition is shown in one unit cell in Figure~\ref{Fig_22}. The analysis shows negligible contribution of the higher-order moments so that the optical response of the unit cell is governed by electric dipole (ED) and magnetic dipole (MD) modes. Thus, the resonances in the reflectance spectra shown in Figures~\ref{Fig_2}d-f are attributed to the strong coupling between the directly reflected light with the ED and MD modes excited inside the PCM nanopillars. These high-Q resonances result in high-saturation colors, which in the case of A-Sb$_2$S$_3$ lie outside the sRGB color space.

To add the polarization-sensitivity to our color-switching approach, from now on, we consider elliptical nanopillars in asymmetric unit cells with different periodicities in the x- and y-directions, i.e. $p_x$ and $p_y$, Figure~\ref{Fig_1}b. By varying $p_x$ and $p_y$ with a fixed ratio with respect to the major and minor axes of the nanopillars (i.e., $d_{x,y}=\alpha \, p_{x,y}$), we generate the color palettes shown in Figures~\ref{Fig_3}a,d,g, for the case of Sb$_2$S$_3$ ($p_{x,y}$ range from 310~nm to 470~nm with 40-nm increments), Sb$_2$Se$_3$ ($p_{x,y}$ range from 200~nm to 400~nm with 50-nm increments), and GeSe$_3$ ($p_{x,y}$ range from 270~nm to 430~nm with 40-nm increments), respectively (see Figures~S4-S6 in the Supporting Information for full color palettes). In each figure, the top (bottom) panels show the colors generated by the x-polarized (y-polarized) incident white light for amorphous (left panels) and crystalline (right panels) cases. While Sb$_2$S$_3$ and GeSe$_3$ metasurfaces can generate a full palette considering both amorphous and crystalline phases (see Figures~\ref{Fig_3}a,g), respectively, Sb$_2$Se$_3$ metasurfaces cannot generate bluish colors (see Figure~\ref{Fig_3}d). This stems from the high optical loss of Sb$_2$Se$_3$ within the blue range of the visible wavelengths (see Figures~\ref{Fig_1}e,f). It is also clear that the y-polarization palettes can be obtained by transposing the x-polarization palette, i.e., replacing each (j,i) element with corresponding (i,j) element. However, this is not the case for amorphous and crystalline palettes in Figures~\ref{Fig_3}a,d,g since the crystalline palettes contain completely different colors from those in the amorphous palettes. This shows the advantage of using PCMs as the number of colors in the phase-transition-based color-switching approach is twice as many as those in the polarization-based approach.

To analyze the effect of polarization-sensitivity in both amorphous and crystalline cases on the reflected colors, we select five metasurfaces for each PCM with geometrical parameters in the dashed boxes in Figures~\ref{Fig_3}a,d,g, and plot the corresponding simulated reflectance spectra with their hue and saturation values in the inset in Figures~\ref{Fig_3}b,e,h, respectively. It is seen that by increasing $p_y$ in each box, the central wavelength of the reflectance resonances does not experience a considerable shift for the x-polarization (see the top panels in Figures~\ref{Fig_3}b,e,h). This leads to almost unchanged hue values for the corresponding colors, which in turn results in a limited trajectory in the corresponding color gamuts shown in the top panels of Figures~\ref{Fig_3}c,f,i in which black circles (white squares) represent the colors in amorphous (crystalline) phase. In contrast, it is observed that increasing $p_y$ results in a tangible redshift in the reflectance spectra for the y-polarization for all PCMs (see the bottom panels in Figures~\ref{Fig_3}b,e,h). This redshift results in a relatively large hue change in all cases, except C-Sb$_2$Se$_3$, as the corresponding color gamuts in the bottom panels of Figures~\ref{Fig_3}c,f,i demonstrate. In the Supporting Information Note~II, we show that by continuously varying the incident polarization angle ($\varphi$) one can enable dynamic color tuning (See Figure~S7). Moreover, the effect of the angle of incident light on the reflected colors is analyzed in the Supporting Information Note~III (See Figure~S8).

To demonstrate the application of our color-switching approach to practical image-imprinting scenarios, we encrypt four images into an array of metasurfaces (pixels) each consisting of elliptic Sb$_2$Se$_3$ nanopillars as shown in Figure~\ref{Fig_5}a. The polarization-based color-switching mechanism is responsible for switching one image seen under x-polarized light to a totally different image for the case of y-polarized illumination. Meanwhile, using the unique feature of the Sb$_2$Se$_3$ metasurfaces in mapping a palette of different colors in the amorphous state into a palette of relatively similar colors in the crystalline state (see Figure~\ref{Fig_3}d), we demonstrate the encoding and decoding of four different images into the C-Sb$_2$Se$_3$ (OFF-state) and A-Sb$_2$Se$_3$ (ON-state), respectively. These capabilities can be used in many applications such as information coding, cryptography, high-density optical data storage, security encryption, and 3D displays. Moreover, same images in Figure~\ref{Fig_5}a are imprinted using Sb$_2$S$_3$ metasurfaces in amorphous phase as shown in Figure~\ref{Fig_5}b. Interestingly, by switching to crystalline phase, the logos (symbols) change to symbols (logos). This is for the first time to the best of our knowledge that two different images can be encrypted in one metasurfaces based on phase-change of the constituent materials. The geometrical parameters of the elliptic nanopillars used to generate each pixel of the images shown in Figures~\ref{Fig_5}a,b are reported in Figures~S10c, S11c and S12a,b in the Supporting Information. More examples of encrypting four images into an array of metasurfaces made of GeSe$_3$ nanopillars are represented in Figures~S10a,b and S11a,b in the Supporting Information in that by switching the GeSe$_3$ phase from amorphous to crystalline, two images with the same image contents but totally different colorations are achieved. The phase-transition-based color-switching capability of Sb$_2$S$_3$ can be used for switching off some parts of an image while maintaining the colors of remaining parts. As an interesting example, the image of Cheshire The Cat is generated by A-Sb$_2$S$_3$ metasurfaces illuminated with x-polarization white light as shown in Figure~\ref{Fig_5}c. Upon phase-transition to the crystalline phase, all parts of the body vanish, but the grinning and eyes remain. The geometrical parameters of the metasurfaces used for producing Figure~\ref{Fig_5}c are given in Figure~S13 in the Supporting Information.

\begin{figure}[t]
\centering
\includegraphics[width=1\linewidth, trim={0cm 0cm 0cm 0cm},clip]{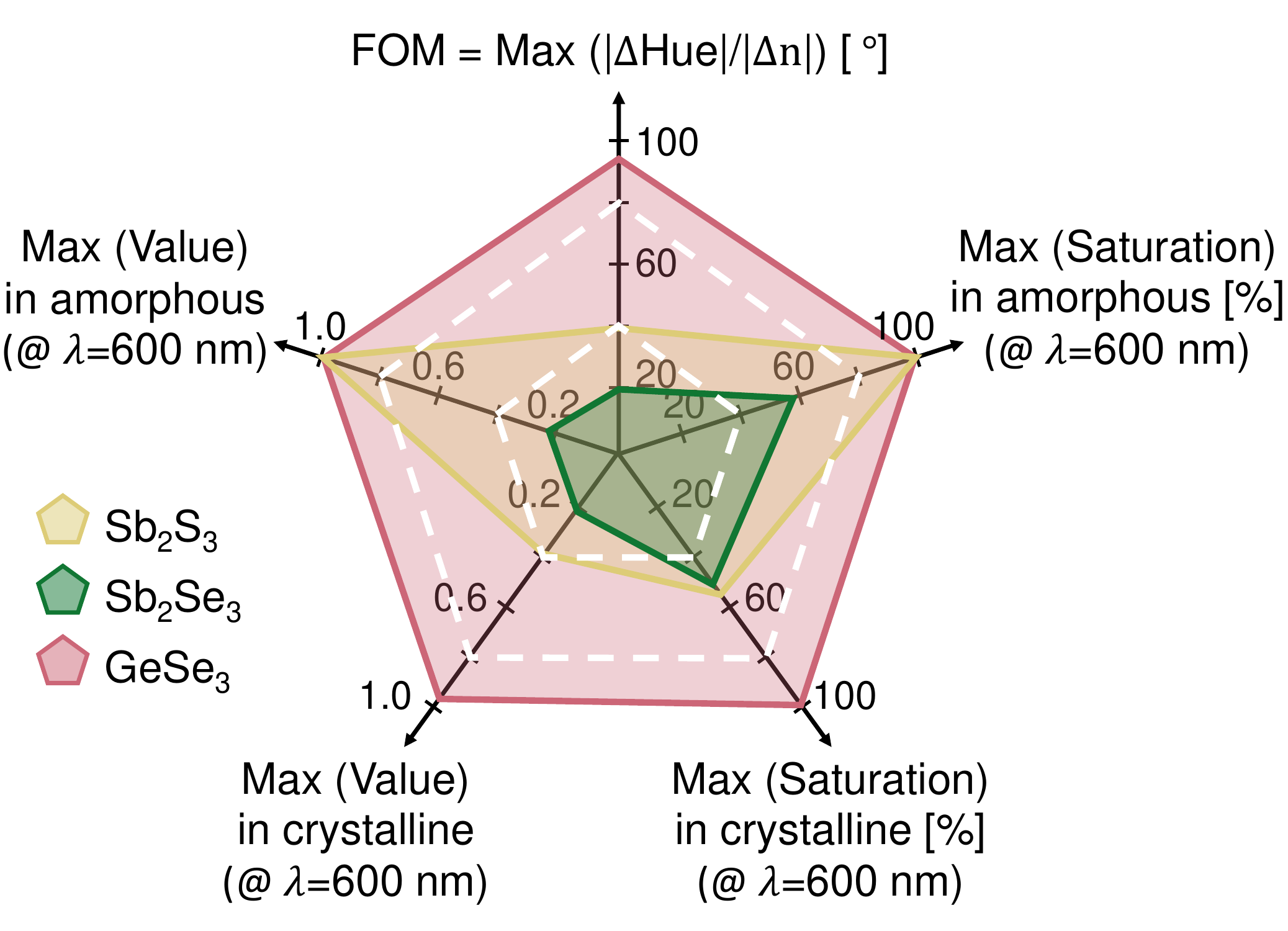}
\caption{\textbf{Comparison of low-loss PCMs for color switching applications.} A spider chart that compares Sb$_2$S$_3$, Sb$_2$Se$_3$ and GeSe$_3$ in terms of FOM (defined as the maximum (max) of $|\Delta \textrm{Hue}|/|\Delta n|$ in which $\Delta n = n_\textrm{A}(\lambda_\textrm{A}) - n_\textrm{C}(\lambda_\textrm{C})$), maximum saturation and maximum value (i.e. the reflectance value at the resonance peak) in amorphous and crystalline phases at $\lambda_\textrm{A} =600$~nm.}
\label{Fig_6}
\end{figure}

In order to demonstrate the potential of our approach in dynamic color printing applications, we re-print a painting of flowers and an image of a Persian tile (the original images are shown in the left panels in Figure~\ref{Fig_5}d and Figure~\ref{Fig_5}e, respectively) using 81 different colors generated by elliptic nanopillars made of GeSe$_3$ and Sb$_2$S$_3$, respectively. The color palettes and corresponding CIE diagrams of these 81 colors are reported in the Supporting Information Figures~S4 and S6. As shown in Figures~\ref{Fig_5}d,e, one can utilize phase-transition- and polarization-based color-switching mechanisms to change the colors of the background and objects, shadowing effects, and contents of the original images at will. The generated four images in Figures~\ref{Fig_5}d and e display different times of the day and different patterns, respectively, achieved using a single reconfigurable metasurface for each pixel in each case.

To provide a comparison between Sb$_2$S$_3$, Sb$_2$Se$_3$, and GeSe$_3$ metasurfaces for color switching applications, a spider chart is shown in Figure~\ref{Fig_6}. The figure of merit (FOM) is defined as the maximum variation of the hue over the refractive index change open phase transition between amorphous and crystalline, i.e. $|\Delta \textrm{Hue}|/|\Delta n|$ in which $\Delta n = n_\textrm{A}(\lambda_\textrm{A}) - n_\textrm{C}(\lambda_\textrm{C})$), with $n_\textrm{A}(\lambda_\textrm{A})$ and $n_\textrm{C}(\lambda_\textrm{C})$ being the index of refraction in the amorphous and crystalline phases and at the corresponding resonance wavelengths ($\lambda_\textrm{A}, \lambda_\textrm{C}$), respectively. While high FOM is desirable, the saturation and value (i.e. the reflectance value at the resonance peak) of the colors in both amorphous and crystalline phases should be as high as possible. Considering all these performance measures, GeSe$_3$ demonstrates superior properties over Sb$_2$S$_3$ and Sb$_2$Se$_3$ when switching from a color associated with a reflectance spectrum with a resonance peak at $\lambda = 600$~nm (chosen as the middle wavelength in the visible range from 400~nm to 800~nm) in the amorphous phase, to another color in the crystalline phase.

\section*{Conclusion}

In summary, we demonstrated a new platform for generating high-efficiency, high-saturation, and wide-gamut switchable structural colors by employing PCM-based metasurfaces made of low-loss and wide-bandgap Sb$_2$S$_3$, Sb$_2$Se$_3$, and GeSe$_3$ nanopillars. Upon the nonvolatile phase-transition of the constituent PCMs, the generated color in the amorphous phase switches to a distinctive stable color in the crystalline phase. In addition, the properly designed asymmetric characteristics of elliptic nanopillars enable polarization-based color switching. Combining these two tuning mechanisms, we systematically designed a single-layer metasurface, which can be used to imprint one pixel of an image in which each pixel is capable of producing four different colors. More interestingly, by gradually changing the crystallinity of the constituent PCMs and/or the incident polarization angle, our platform enable realization of multi-color artificial images. We also showed that by selecting the proper PCM, features like ON/OFF switching, color shading, and image switching can be implemented. We believe that this research can route a key step towards commercialization of full-color dynamic displays, information storage, image encryption, and anti-counterfeiting.

\section*{Acknowledgements}

The work was primarily funded by the Office of Naval Research (ONR) (N00014-18-1-2055, Dr. B. Bennett) and by the Air Force Office of Scientific Research MURI program.

\section*{Disclosures} The authors declare no conflicts of interest.

%


\clearpage
\newpage


\begin{thebibliography}{84}%
\makeatletter
\providecommand \@ifxundefined [1]{%
 \@ifx{#1\undefined}
}%
\providecommand \@ifnum [1]{%
 \ifnum #1\expandafter \@firstoftwo
 \else \expandafter \@secondoftwo
 \fi
}%
\providecommand \@ifx [1]{%
 \ifx #1\expandafter \@firstoftwo
 \else \expandafter \@secondoftwo
 \fi
}%
\providecommand \natexlab [1]{#1}%
\providecommand \enquote  [1]{``#1''}%
\providecommand \bibnamefont  [1]{#1}%
\providecommand \bibfnamefont [1]{#1}%
\providecommand \citenamefont [1]{#1}%
\providecommand \href@noop [0]{\@secondoftwo}%
\providecommand \href [0]{\begingroup \@sanitize@url \@href}%
\providecommand \@href[1]{\@@startlink{#1}\@@href}%
\providecommand \@@href[1]{\endgroup#1\@@endlink}%
\providecommand \@sanitize@url [0]{\catcode `\\12\catcode `\$12\catcode
  `\&12\catcode `\#12\catcode `\^12\catcode `\_12\catcode `\%12\relax}%
\providecommand \@@startlink[1]{}%
\providecommand \@@endlink[0]{}%
\providecommand \url  [0]{\begingroup\@sanitize@url \@url }%
\providecommand \@url [1]{\endgroup\@href {#1}{\urlprefix }}%
\providecommand \urlprefix  [0]{URL }%
\providecommand \Eprint [0]{\href }%
\providecommand \doibase [0]{https://doi.org/}%
\providecommand \selectlanguage [0]{\@gobble}%
\providecommand \bibinfo  [0]{\@secondoftwo}%
\providecommand \bibfield  [0]{\@secondoftwo}%
\providecommand \translation [1]{[#1]}%
\providecommand \BibitemOpen [0]{}%
\providecommand \bibitemStop [0]{}%
\providecommand \bibitemNoStop [0]{.\EOS\space}%
\providecommand \EOS [0]{\spacefactor3000\relax}%
\providecommand \BibitemShut  [1]{\csname bibitem#1\endcsname}%
\let\auto@bib@innerbib\@empty
\bibitem [{\citenamefont {Daqiqeh~Rezaei}\ \emph {et~al.}(2020)\citenamefont
  {Daqiqeh~Rezaei}, \citenamefont {Dong}, \citenamefont {You En~Chan},
  \citenamefont {Trisno}, \citenamefont {Ng}, \citenamefont {Ruan},
  \citenamefont {Qiu}, \citenamefont {Mortensen},\ and\ \citenamefont
  {Yang}}]{daqiqeh2020nanophotonic}%
  \BibitemOpen
  \bibfield  {author} {\bibinfo {author} {\bibfnamefont {S.}~\bibnamefont
  {Daqiqeh~Rezaei}}, \bibinfo {author} {\bibfnamefont {Z.}~\bibnamefont
  {Dong}}, \bibinfo {author} {\bibfnamefont {J.}~\bibnamefont {You En~Chan}},
  \bibinfo {author} {\bibfnamefont {J.}~\bibnamefont {Trisno}}, \bibinfo
  {author} {\bibfnamefont {R.~J.~H.}\ \bibnamefont {Ng}}, \bibinfo {author}
  {\bibfnamefont {Q.}~\bibnamefont {Ruan}}, \bibinfo {author} {\bibfnamefont
  {C.-W.}\ \bibnamefont {Qiu}}, \bibinfo {author} {\bibfnamefont {N.~A.}\
  \bibnamefont {Mortensen}},\ and\ \bibinfo {author} {\bibfnamefont {J.~K.}\
  \bibnamefont {Yang}},\ }\bibfield  {title} {\bibinfo {title} {Nanophotonic
  structural colors},\ }\href@noop {} {\bibfield  {journal} {\bibinfo
  {journal} {ACS Photonics}\ } (\bibinfo {year} {2020})}\BibitemShut {NoStop}%
\bibitem [{\citenamefont {Vukusic}\ \emph {et~al.}(1999)\citenamefont
  {Vukusic}, \citenamefont {Sambles}, \citenamefont {Lawrence},\ and\
  \citenamefont {Wootton}}]{vukusic1999quantified}%
  \BibitemOpen
  \bibfield  {author} {\bibinfo {author} {\bibfnamefont {P.}~\bibnamefont
  {Vukusic}}, \bibinfo {author} {\bibfnamefont {J.}~\bibnamefont {Sambles}},
  \bibinfo {author} {\bibfnamefont {C.}~\bibnamefont {Lawrence}},\ and\
  \bibinfo {author} {\bibfnamefont {R.}~\bibnamefont {Wootton}},\ }\bibfield
  {title} {\bibinfo {title} {Quantified interference and diffraction in single
  morpho butterfly scales},\ }\href@noop {} {\bibfield  {journal} {\bibinfo
  {journal} {Proceedings of the Royal Society of London. Series B: Biological
  Sciences}\ }\textbf {\bibinfo {volume} {266}},\ \bibinfo {pages} {1403}
  (\bibinfo {year} {1999})}\BibitemShut {NoStop}%
\bibitem [{\citenamefont {Al{\`u}}\ and\ \citenamefont
  {Engheta}(2005)}]{alu2005achieving}%
  \BibitemOpen
  \bibfield  {author} {\bibinfo {author} {\bibfnamefont {A.}~\bibnamefont
  {Al{\`u}}}\ and\ \bibinfo {author} {\bibfnamefont {N.}~\bibnamefont
  {Engheta}},\ }\bibfield  {title} {\bibinfo {title} {Achieving transparency
  with plasmonic and metamaterial coatings},\ }\href@noop {} {\bibfield
  {journal} {\bibinfo  {journal} {Physical Review E}\ }\textbf {\bibinfo
  {volume} {72}},\ \bibinfo {pages} {016623} (\bibinfo {year}
  {2005})}\BibitemShut {NoStop}%
\bibitem [{\citenamefont {Krasnok}\ \emph {et~al.}(2012)\citenamefont
  {Krasnok}, \citenamefont {Miroshnichenko}, \citenamefont {Belov},\ and\
  \citenamefont {Kivshar}}]{krasnok2012all}%
  \BibitemOpen
  \bibfield  {author} {\bibinfo {author} {\bibfnamefont {A.~E.}\ \bibnamefont
  {Krasnok}}, \bibinfo {author} {\bibfnamefont {A.~E.}\ \bibnamefont
  {Miroshnichenko}}, \bibinfo {author} {\bibfnamefont {P.~A.}\ \bibnamefont
  {Belov}},\ and\ \bibinfo {author} {\bibfnamefont {Y.~S.}\ \bibnamefont
  {Kivshar}},\ }\bibfield  {title} {\bibinfo {title} {All-dielectric optical
  nanoantennas},\ }\href@noop {} {\bibfield  {journal} {\bibinfo  {journal}
  {Optics Express}\ }\textbf {\bibinfo {volume} {20}},\ \bibinfo {pages}
  {20599} (\bibinfo {year} {2012})}\BibitemShut {NoStop}%
\bibitem [{\citenamefont {Yu}\ and\ \citenamefont
  {Capasso}(2014)}]{yu2014flat}%
  \BibitemOpen
  \bibfield  {author} {\bibinfo {author} {\bibfnamefont {N.}~\bibnamefont
  {Yu}}\ and\ \bibinfo {author} {\bibfnamefont {F.}~\bibnamefont {Capasso}},\
  }\bibfield  {title} {\bibinfo {title} {Flat optics with designer
  metasurfaces},\ }\href@noop {} {\bibfield  {journal} {\bibinfo  {journal}
  {Nature materials}\ }\textbf {\bibinfo {volume} {13}},\ \bibinfo {pages}
  {139} (\bibinfo {year} {2014})}\BibitemShut {NoStop}%
\bibitem [{\citenamefont {Miroshnichenko}\ \emph {et~al.}(2010)\citenamefont
  {Miroshnichenko}, \citenamefont {Flach},\ and\ \citenamefont
  {Kivshar}}]{miroshnichenko2010fano}%
  \BibitemOpen
  \bibfield  {author} {\bibinfo {author} {\bibfnamefont {A.~E.}\ \bibnamefont
  {Miroshnichenko}}, \bibinfo {author} {\bibfnamefont {S.}~\bibnamefont
  {Flach}},\ and\ \bibinfo {author} {\bibfnamefont {Y.~S.}\ \bibnamefont
  {Kivshar}},\ }\bibfield  {title} {\bibinfo {title} {Fano resonances in
  nanoscale structures},\ }\href@noop {} {\bibfield  {journal} {\bibinfo
  {journal} {Reviews of Modern Physics}\ }\textbf {\bibinfo {volume} {82}},\
  \bibinfo {pages} {2257} (\bibinfo {year} {2010})}\BibitemShut {NoStop}%
\bibitem [{\citenamefont {Decker}\ \emph {et~al.}(2015)\citenamefont {Decker},
  \citenamefont {Staude}, \citenamefont {Falkner}, \citenamefont {Dominguez},
  \citenamefont {Neshev}, \citenamefont {Brener}, \citenamefont {Pertsch},\
  and\ \citenamefont {Kivshar}}]{decker2015high}%
  \BibitemOpen
  \bibfield  {author} {\bibinfo {author} {\bibfnamefont {M.}~\bibnamefont
  {Decker}}, \bibinfo {author} {\bibfnamefont {I.}~\bibnamefont {Staude}},
  \bibinfo {author} {\bibfnamefont {M.}~\bibnamefont {Falkner}}, \bibinfo
  {author} {\bibfnamefont {J.}~\bibnamefont {Dominguez}}, \bibinfo {author}
  {\bibfnamefont {D.~N.}\ \bibnamefont {Neshev}}, \bibinfo {author}
  {\bibfnamefont {I.}~\bibnamefont {Brener}}, \bibinfo {author} {\bibfnamefont
  {T.}~\bibnamefont {Pertsch}},\ and\ \bibinfo {author} {\bibfnamefont {Y.~S.}\
  \bibnamefont {Kivshar}},\ }\bibfield  {title} {\bibinfo {title}
  {High-efficiency dielectric huygens’ surfaces},\ }\href@noop {} {\bibfield
  {journal} {\bibinfo  {journal} {Advanced Optical Materials}\ }\textbf
  {\bibinfo {volume} {3}},\ \bibinfo {pages} {813} (\bibinfo {year}
  {2015})}\BibitemShut {NoStop}%
\bibitem [{\citenamefont {AbdollahRamezani}\ \emph {et~al.}(2015)\citenamefont
  {AbdollahRamezani}, \citenamefont {Arik}, \citenamefont {Khavasi},\ and\
  \citenamefont {Kavehvash}}]{abdollahramezani2015analog}%
  \BibitemOpen
  \bibfield  {author} {\bibinfo {author} {\bibfnamefont {S.}~\bibnamefont
  {AbdollahRamezani}}, \bibinfo {author} {\bibfnamefont {K.}~\bibnamefont
  {Arik}}, \bibinfo {author} {\bibfnamefont {A.}~\bibnamefont {Khavasi}},\ and\
  \bibinfo {author} {\bibfnamefont {Z.}~\bibnamefont {Kavehvash}},\ }\bibfield
  {title} {\bibinfo {title} {Analog computing using graphene-based metalines},\
  }\href@noop {} {\bibfield  {journal} {\bibinfo  {journal} {Optics letters}\
  }\textbf {\bibinfo {volume} {40}},\ \bibinfo {pages} {5239} (\bibinfo {year}
  {2015})}\BibitemShut {NoStop}%
\bibitem [{\citenamefont {Kuznetsov}\ \emph {et~al.}(2016)\citenamefont
  {Kuznetsov}, \citenamefont {Miroshnichenko}, \citenamefont {Brongersma},
  \citenamefont {Kivshar},\ and\ \citenamefont
  {Luk’yanchuk}}]{kuznetsov2016optically}%
  \BibitemOpen
  \bibfield  {author} {\bibinfo {author} {\bibfnamefont {A.~I.}\ \bibnamefont
  {Kuznetsov}}, \bibinfo {author} {\bibfnamefont {A.~E.}\ \bibnamefont
  {Miroshnichenko}}, \bibinfo {author} {\bibfnamefont {M.~L.}\ \bibnamefont
  {Brongersma}}, \bibinfo {author} {\bibfnamefont {Y.~S.}\ \bibnamefont
  {Kivshar}},\ and\ \bibinfo {author} {\bibfnamefont {B.}~\bibnamefont
  {Luk’yanchuk}},\ }\bibfield  {title} {\bibinfo {title} {Optically resonant
  dielectric nanostructures},\ }\href@noop {} {\bibfield  {journal} {\bibinfo
  {journal} {Science}\ }\textbf {\bibinfo {volume} {354}} (\bibinfo {year}
  {2016})}\BibitemShut {NoStop}%
\bibitem [{\citenamefont {Hemmatyar}\ \emph {et~al.}(2017)\citenamefont
  {Hemmatyar}, \citenamefont {Rahmani}, \citenamefont {Bagheri},\ and\
  \citenamefont {Khavasi}}]{hemmatyar2017phase}%
  \BibitemOpen
  \bibfield  {author} {\bibinfo {author} {\bibfnamefont {O.}~\bibnamefont
  {Hemmatyar}}, \bibinfo {author} {\bibfnamefont {B.}~\bibnamefont {Rahmani}},
  \bibinfo {author} {\bibfnamefont {A.}~\bibnamefont {Bagheri}},\ and\ \bibinfo
  {author} {\bibfnamefont {A.}~\bibnamefont {Khavasi}},\ }\bibfield  {title}
  {\bibinfo {title} {Phase resonance tuning and multi-band absorption via
  graphene-covered compound metallic gratings},\ }\href@noop {} {\bibfield
  {journal} {\bibinfo  {journal} {IEEE Journal of Quantum Electronics}\
  }\textbf {\bibinfo {volume} {53}},\ \bibinfo {pages} {1} (\bibinfo {year}
  {2017})}\BibitemShut {NoStop}%
\bibitem [{\citenamefont {Hu}\ \emph {et~al.}(2020)\citenamefont {Hu},
  \citenamefont {Krasnok}, \citenamefont {Mazor}, \citenamefont {Qiu},\ and\
  \citenamefont {Al{\`u}}}]{hu2020moire}%
  \BibitemOpen
  \bibfield  {author} {\bibinfo {author} {\bibfnamefont {G.}~\bibnamefont
  {Hu}}, \bibinfo {author} {\bibfnamefont {A.}~\bibnamefont {Krasnok}},
  \bibinfo {author} {\bibfnamefont {Y.}~\bibnamefont {Mazor}}, \bibinfo
  {author} {\bibfnamefont {C.-W.}\ \bibnamefont {Qiu}},\ and\ \bibinfo {author}
  {\bibfnamefont {A.}~\bibnamefont {Al{\`u}}},\ }\bibfield  {title} {\bibinfo
  {title} {Moir{\'e} hyperbolic metasurfaces},\ }\href@noop {} {\bibfield
  {journal} {\bibinfo  {journal} {Nano Letters}\ }\textbf {\bibinfo {volume}
  {20}},\ \bibinfo {pages} {3217} (\bibinfo {year} {2020})}\BibitemShut
  {NoStop}%
\bibitem [{\citenamefont {Arik}\ \emph {et~al.}(2020)\citenamefont {Arik},
  \citenamefont {Hemmatyar},\ and\ \citenamefont {Kavehvash}}]{arik2020beam}%
  \BibitemOpen
  \bibfield  {author} {\bibinfo {author} {\bibfnamefont {K.}~\bibnamefont
  {Arik}}, \bibinfo {author} {\bibfnamefont {O.}~\bibnamefont {Hemmatyar}},\
  and\ \bibinfo {author} {\bibfnamefont {Z.}~\bibnamefont {Kavehvash}},\
  }\bibfield  {title} {\bibinfo {title} {Beam manipulation by hybrid
  plasmonic-dielectric metasurfaces},\ }\href@noop {} {\bibfield  {journal}
  {\bibinfo  {journal} {Plasmonics}\ }\textbf {\bibinfo {volume} {15}},\
  \bibinfo {pages} {639} (\bibinfo {year} {2020})}\BibitemShut {NoStop}%
\bibitem [{\citenamefont {Hemmatyar}\ \emph
  {et~al.}(2020{\natexlab{a}})\citenamefont {Hemmatyar}, \citenamefont
  {Abbassi}, \citenamefont {Rahmani}, \citenamefont {Memarian},\ and\
  \citenamefont {Mehrany}}]{hemmatyar2020wide}%
  \BibitemOpen
  \bibfield  {author} {\bibinfo {author} {\bibfnamefont {O.}~\bibnamefont
  {Hemmatyar}}, \bibinfo {author} {\bibfnamefont {M.~A.}\ \bibnamefont
  {Abbassi}}, \bibinfo {author} {\bibfnamefont {B.}~\bibnamefont {Rahmani}},
  \bibinfo {author} {\bibfnamefont {M.}~\bibnamefont {Memarian}},\ and\
  \bibinfo {author} {\bibfnamefont {K.}~\bibnamefont {Mehrany}},\ }\bibfield
  {title} {\bibinfo {title} {Wide-band/angle blazed dual mode metallic groove
  gratings},\ }\href@noop {} {\bibfield  {journal} {\bibinfo  {journal} {IEEE
  Transactions on Antennas and Propagation}\ } (\bibinfo {year}
  {2020}{\natexlab{a}})}\BibitemShut {NoStop}%
\bibitem [{\citenamefont {Abdollahramezani}\ \emph
  {et~al.}(2020{\natexlab{a}})\citenamefont {Abdollahramezani}, \citenamefont
  {Hemmatyar},\ and\ \citenamefont {Adibi}}]{abdollahramezani2020meta}%
  \BibitemOpen
  \bibfield  {author} {\bibinfo {author} {\bibfnamefont {S.}~\bibnamefont
  {Abdollahramezani}}, \bibinfo {author} {\bibfnamefont {O.}~\bibnamefont
  {Hemmatyar}},\ and\ \bibinfo {author} {\bibfnamefont {A.}~\bibnamefont
  {Adibi}},\ }\bibfield  {title} {\bibinfo {title} {Meta-optics for spatial
  optical analog computing},\ }\href@noop {} {\bibfield  {journal} {\bibinfo
  {journal} {Nanophotonics}\ }\textbf {\bibinfo {volume} {9}},\ \bibinfo
  {pages} {4075} (\bibinfo {year} {2020}{\natexlab{a}})}\BibitemShut {NoStop}%
\bibitem [{\citenamefont {Kumar}\ \emph {et~al.}(2012)\citenamefont {Kumar},
  \citenamefont {Duan}, \citenamefont {Hegde}, \citenamefont {Koh},
  \citenamefont {Wei},\ and\ \citenamefont {Yang}}]{kumar2012printing}%
  \BibitemOpen
  \bibfield  {author} {\bibinfo {author} {\bibfnamefont {K.}~\bibnamefont
  {Kumar}}, \bibinfo {author} {\bibfnamefont {H.}~\bibnamefont {Duan}},
  \bibinfo {author} {\bibfnamefont {R.~S.}\ \bibnamefont {Hegde}}, \bibinfo
  {author} {\bibfnamefont {S.~C.}\ \bibnamefont {Koh}}, \bibinfo {author}
  {\bibfnamefont {J.~N.}\ \bibnamefont {Wei}},\ and\ \bibinfo {author}
  {\bibfnamefont {J.~K.}\ \bibnamefont {Yang}},\ }\bibfield  {title} {\bibinfo
  {title} {Printing colour at the optical diffraction limit},\ }\href@noop {}
  {\bibfield  {journal} {\bibinfo  {journal} {Nature nanotechnology}\ }\textbf
  {\bibinfo {volume} {7}},\ \bibinfo {pages} {557} (\bibinfo {year}
  {2012})}\BibitemShut {NoStop}%
\bibitem [{\citenamefont {Tan}\ \emph {et~al.}(2014)\citenamefont {Tan},
  \citenamefont {Zhang}, \citenamefont {Zhu}, \citenamefont {Goh},
  \citenamefont {Wang}, \citenamefont {Kumar}, \citenamefont {Qiu},\ and\
  \citenamefont {Yang}}]{tan2014plasmonic}%
  \BibitemOpen
  \bibfield  {author} {\bibinfo {author} {\bibfnamefont {S.~J.}\ \bibnamefont
  {Tan}}, \bibinfo {author} {\bibfnamefont {L.}~\bibnamefont {Zhang}}, \bibinfo
  {author} {\bibfnamefont {D.}~\bibnamefont {Zhu}}, \bibinfo {author}
  {\bibfnamefont {X.~M.}\ \bibnamefont {Goh}}, \bibinfo {author} {\bibfnamefont
  {Y.~M.}\ \bibnamefont {Wang}}, \bibinfo {author} {\bibfnamefont
  {K.}~\bibnamefont {Kumar}}, \bibinfo {author} {\bibfnamefont {C.-W.}\
  \bibnamefont {Qiu}},\ and\ \bibinfo {author} {\bibfnamefont {J.~K.}\
  \bibnamefont {Yang}},\ }\bibfield  {title} {\bibinfo {title} {Plasmonic color
  palettes for photorealistic printing with aluminum nanostructures},\
  }\href@noop {} {\bibfield  {journal} {\bibinfo  {journal} {Nano letters}\
  }\textbf {\bibinfo {volume} {14}},\ \bibinfo {pages} {4023} (\bibinfo {year}
  {2014})}\BibitemShut {NoStop}%
\bibitem [{\citenamefont {Kristensen}\ \emph {et~al.}(2016)\citenamefont
  {Kristensen}, \citenamefont {Yang}, \citenamefont {Bozhevolnyi},
  \citenamefont {Link}, \citenamefont {Nordlander}, \citenamefont {Halas},\
  and\ \citenamefont {Mortensen}}]{kristensen2016plasmonic}%
  \BibitemOpen
  \bibfield  {author} {\bibinfo {author} {\bibfnamefont {A.}~\bibnamefont
  {Kristensen}}, \bibinfo {author} {\bibfnamefont {J.~K.}\ \bibnamefont
  {Yang}}, \bibinfo {author} {\bibfnamefont {S.~I.}\ \bibnamefont
  {Bozhevolnyi}}, \bibinfo {author} {\bibfnamefont {S.}~\bibnamefont {Link}},
  \bibinfo {author} {\bibfnamefont {P.}~\bibnamefont {Nordlander}}, \bibinfo
  {author} {\bibfnamefont {N.~J.}\ \bibnamefont {Halas}},\ and\ \bibinfo
  {author} {\bibfnamefont {N.~A.}\ \bibnamefont {Mortensen}},\ }\bibfield
  {title} {\bibinfo {title} {Plasmonic colour generation},\ }\href@noop {}
  {\bibfield  {journal} {\bibinfo  {journal} {Nature Reviews Materials}\
  }\textbf {\bibinfo {volume} {2}},\ \bibinfo {pages} {1} (\bibinfo {year}
  {2016})}\BibitemShut {NoStop}%
\bibitem [{\citenamefont {Rezaei}\ \emph {et~al.}(2019)\citenamefont {Rezaei},
  \citenamefont {Hong~Ng}, \citenamefont {Dong}, \citenamefont {Ho},
  \citenamefont {Koay}, \citenamefont {Ramakrishna},\ and\ \citenamefont
  {Yang}}]{rezaei2019wide}%
  \BibitemOpen
  \bibfield  {author} {\bibinfo {author} {\bibfnamefont {S.~D.}\ \bibnamefont
  {Rezaei}}, \bibinfo {author} {\bibfnamefont {R.~J.}\ \bibnamefont {Hong~Ng}},
  \bibinfo {author} {\bibfnamefont {Z.}~\bibnamefont {Dong}}, \bibinfo {author}
  {\bibfnamefont {J.}~\bibnamefont {Ho}}, \bibinfo {author} {\bibfnamefont
  {E.~H.}\ \bibnamefont {Koay}}, \bibinfo {author} {\bibfnamefont
  {S.}~\bibnamefont {Ramakrishna}},\ and\ \bibinfo {author} {\bibfnamefont
  {J.~K.}\ \bibnamefont {Yang}},\ }\bibfield  {title} {\bibinfo {title}
  {Wide-gamut plasmonic color palettes with constant subwavelength
  resolution},\ }\href@noop {} {\bibfield  {journal} {\bibinfo  {journal} {ACS
  nano}\ }\textbf {\bibinfo {volume} {13}},\ \bibinfo {pages} {3580} (\bibinfo
  {year} {2019})}\BibitemShut {NoStop}%
\bibitem [{\citenamefont {Shen}\ \emph {et~al.}(2015)\citenamefont {Shen},
  \citenamefont {Rinnerbauer}, \citenamefont {Wang}, \citenamefont {Stelmakh},
  \citenamefont {Joannopoulos},\ and\ \citenamefont
  {Soljacic}}]{shen2015structural}%
  \BibitemOpen
  \bibfield  {author} {\bibinfo {author} {\bibfnamefont {Y.}~\bibnamefont
  {Shen}}, \bibinfo {author} {\bibfnamefont {V.}~\bibnamefont {Rinnerbauer}},
  \bibinfo {author} {\bibfnamefont {I.}~\bibnamefont {Wang}}, \bibinfo {author}
  {\bibfnamefont {V.}~\bibnamefont {Stelmakh}}, \bibinfo {author}
  {\bibfnamefont {J.~D.}\ \bibnamefont {Joannopoulos}},\ and\ \bibinfo {author}
  {\bibfnamefont {M.}~\bibnamefont {Soljacic}},\ }\bibfield  {title} {\bibinfo
  {title} {Structural colors from fano resonances},\ }\href@noop {} {\bibfield
  {journal} {\bibinfo  {journal} {Acs Photonics}\ }\textbf {\bibinfo {volume}
  {2}},\ \bibinfo {pages} {27} (\bibinfo {year} {2015})}\BibitemShut {NoStop}%
\bibitem [{\citenamefont {Sun}\ \emph {et~al.}(2017)\citenamefont {Sun},
  \citenamefont {Zhou}, \citenamefont {Zhang}, \citenamefont {Gao},
  \citenamefont {Duan}, \citenamefont {Xiao},\ and\ \citenamefont
  {Song}}]{sun2017all}%
  \BibitemOpen
  \bibfield  {author} {\bibinfo {author} {\bibfnamefont {S.}~\bibnamefont
  {Sun}}, \bibinfo {author} {\bibfnamefont {Z.}~\bibnamefont {Zhou}}, \bibinfo
  {author} {\bibfnamefont {C.}~\bibnamefont {Zhang}}, \bibinfo {author}
  {\bibfnamefont {Y.}~\bibnamefont {Gao}}, \bibinfo {author} {\bibfnamefont
  {Z.}~\bibnamefont {Duan}}, \bibinfo {author} {\bibfnamefont {S.}~\bibnamefont
  {Xiao}},\ and\ \bibinfo {author} {\bibfnamefont {Q.}~\bibnamefont {Song}},\
  }\bibfield  {title} {\bibinfo {title} {All-dielectric full-color printing
  with tio2 metasurfaces},\ }\href@noop {} {\bibfield  {journal} {\bibinfo
  {journal} {ACS nano}\ }\textbf {\bibinfo {volume} {11}},\ \bibinfo {pages}
  {4445} (\bibinfo {year} {2017})}\BibitemShut {NoStop}%
\bibitem [{\citenamefont {Zhu}\ \emph {et~al.}(2017)\citenamefont {Zhu},
  \citenamefont {Yan}, \citenamefont {Levy}, \citenamefont {Mortensen},\ and\
  \citenamefont {Kristensen}}]{zhu2017resonant}%
  \BibitemOpen
  \bibfield  {author} {\bibinfo {author} {\bibfnamefont {X.}~\bibnamefont
  {Zhu}}, \bibinfo {author} {\bibfnamefont {W.}~\bibnamefont {Yan}}, \bibinfo
  {author} {\bibfnamefont {U.}~\bibnamefont {Levy}}, \bibinfo {author}
  {\bibfnamefont {N.~A.}\ \bibnamefont {Mortensen}},\ and\ \bibinfo {author}
  {\bibfnamefont {A.}~\bibnamefont {Kristensen}},\ }\bibfield  {title}
  {\bibinfo {title} {Resonant laser printing of structural colors on high-index
  dielectric metasurfaces},\ }\href@noop {} {\bibfield  {journal} {\bibinfo
  {journal} {Science advances}\ }\textbf {\bibinfo {volume} {3}},\ \bibinfo
  {pages} {e1602487} (\bibinfo {year} {2017})}\BibitemShut {NoStop}%
\bibitem [{\citenamefont {Dong}\ \emph {et~al.}(2017)\citenamefont {Dong},
  \citenamefont {Ho}, \citenamefont {Yu}, \citenamefont {Fu}, \citenamefont
  {Paniagua-Dominguez}, \citenamefont {Wang}, \citenamefont {Kuznetsov},\ and\
  \citenamefont {Yang}}]{dong2017printing}%
  \BibitemOpen
  \bibfield  {author} {\bibinfo {author} {\bibfnamefont {Z.}~\bibnamefont
  {Dong}}, \bibinfo {author} {\bibfnamefont {J.}~\bibnamefont {Ho}}, \bibinfo
  {author} {\bibfnamefont {Y.~F.}\ \bibnamefont {Yu}}, \bibinfo {author}
  {\bibfnamefont {Y.~H.}\ \bibnamefont {Fu}}, \bibinfo {author} {\bibfnamefont
  {R.}~\bibnamefont {Paniagua-Dominguez}}, \bibinfo {author} {\bibfnamefont
  {S.}~\bibnamefont {Wang}}, \bibinfo {author} {\bibfnamefont {A.~I.}\
  \bibnamefont {Kuznetsov}},\ and\ \bibinfo {author} {\bibfnamefont {J.~K.}\
  \bibnamefont {Yang}},\ }\bibfield  {title} {\bibinfo {title} {Printing beyond
  srgb color gamut by mimicking silicon nanostructures in free-space},\
  }\href@noop {} {\bibfield  {journal} {\bibinfo  {journal} {Nano letters}\
  }\textbf {\bibinfo {volume} {17}},\ \bibinfo {pages} {7620} (\bibinfo {year}
  {2017})}\BibitemShut {NoStop}%
\bibitem [{\citenamefont {Jin}\ \emph {et~al.}(2018)\citenamefont {Jin},
  \citenamefont {Dong}, \citenamefont {Mei}, \citenamefont {Yu}, \citenamefont
  {Wei}, \citenamefont {Pan}, \citenamefont {Rezaei}, \citenamefont {Li},
  \citenamefont {Kuznetsov}, \citenamefont {Kivshar} \emph
  {et~al.}}]{jin2018noninterleaved}%
  \BibitemOpen
  \bibfield  {author} {\bibinfo {author} {\bibfnamefont {L.}~\bibnamefont
  {Jin}}, \bibinfo {author} {\bibfnamefont {Z.}~\bibnamefont {Dong}}, \bibinfo
  {author} {\bibfnamefont {S.}~\bibnamefont {Mei}}, \bibinfo {author}
  {\bibfnamefont {Y.~F.}\ \bibnamefont {Yu}}, \bibinfo {author} {\bibfnamefont
  {Z.}~\bibnamefont {Wei}}, \bibinfo {author} {\bibfnamefont {Z.}~\bibnamefont
  {Pan}}, \bibinfo {author} {\bibfnamefont {S.~D.}\ \bibnamefont {Rezaei}},
  \bibinfo {author} {\bibfnamefont {X.}~\bibnamefont {Li}}, \bibinfo {author}
  {\bibfnamefont {A.~I.}\ \bibnamefont {Kuznetsov}}, \bibinfo {author}
  {\bibfnamefont {Y.~S.}\ \bibnamefont {Kivshar}}, \emph {et~al.},\ }\bibfield
  {title} {\bibinfo {title} {Noninterleaved metasurface for (26-1) spin-and
  wavelength-encoded holograms},\ }\href@noop {} {\bibfield  {journal}
  {\bibinfo  {journal} {Nano letters}\ }\textbf {\bibinfo {volume} {18}},\
  \bibinfo {pages} {8016} (\bibinfo {year} {2018})}\BibitemShut {NoStop}%
\bibitem [{\citenamefont {Yang}\ \emph {et~al.}(2019)\citenamefont {Yang},
  \citenamefont {Liu}, \citenamefont {Li}, \citenamefont {Cheng}, \citenamefont
  {Choi}, \citenamefont {Chen},\ and\ \citenamefont
  {Tian}}]{yang2019ultrahighly}%
  \BibitemOpen
  \bibfield  {author} {\bibinfo {author} {\bibfnamefont {B.}~\bibnamefont
  {Yang}}, \bibinfo {author} {\bibfnamefont {W.}~\bibnamefont {Liu}}, \bibinfo
  {author} {\bibfnamefont {Z.}~\bibnamefont {Li}}, \bibinfo {author}
  {\bibfnamefont {H.}~\bibnamefont {Cheng}}, \bibinfo {author} {\bibfnamefont
  {D.-Y.}\ \bibnamefont {Choi}}, \bibinfo {author} {\bibfnamefont
  {S.}~\bibnamefont {Chen}},\ and\ \bibinfo {author} {\bibfnamefont
  {J.}~\bibnamefont {Tian}},\ }\bibfield  {title} {\bibinfo {title}
  {Ultrahighly saturated structural colors enhanced by multipolar-modulated
  metasurfaces},\ }\href@noop {} {\bibfield  {journal} {\bibinfo  {journal}
  {Nano letters}\ }\textbf {\bibinfo {volume} {19}},\ \bibinfo {pages} {4221}
  (\bibinfo {year} {2019})}\BibitemShut {NoStop}%
\bibitem [{\citenamefont {Hemmatyar}\ \emph {et~al.}(2019)\citenamefont
  {Hemmatyar}, \citenamefont {Abdollahramezani}, \citenamefont {Kiarashinejad},
  \citenamefont {Zandehshahvar},\ and\ \citenamefont
  {Adibi}}]{hemmatyar2019full}%
  \BibitemOpen
  \bibfield  {author} {\bibinfo {author} {\bibfnamefont {O.}~\bibnamefont
  {Hemmatyar}}, \bibinfo {author} {\bibfnamefont {S.}~\bibnamefont
  {Abdollahramezani}}, \bibinfo {author} {\bibfnamefont {Y.}~\bibnamefont
  {Kiarashinejad}}, \bibinfo {author} {\bibfnamefont {M.}~\bibnamefont
  {Zandehshahvar}},\ and\ \bibinfo {author} {\bibfnamefont {A.}~\bibnamefont
  {Adibi}},\ }\bibfield  {title} {\bibinfo {title} {Full color generation with
  fano-type resonant hfo 2 nanopillars designed by a deep-learning approach},\
  }\href@noop {} {\bibfield  {journal} {\bibinfo  {journal} {Nanoscale}\
  }\textbf {\bibinfo {volume} {11}},\ \bibinfo {pages} {21266} (\bibinfo {year}
  {2019})}\BibitemShut {NoStop}%
\bibitem [{\citenamefont {Yang}\ \emph {et~al.}(2020)\citenamefont {Yang},
  \citenamefont {Xiao}, \citenamefont {Song}, \citenamefont {Liu},
  \citenamefont {Wu}, \citenamefont {Wang}, \citenamefont {Yu}, \citenamefont
  {Han},\ and\ \citenamefont {Tsai}}]{yang2020all}%
  \BibitemOpen
  \bibfield  {author} {\bibinfo {author} {\bibfnamefont {W.}~\bibnamefont
  {Yang}}, \bibinfo {author} {\bibfnamefont {S.}~\bibnamefont {Xiao}}, \bibinfo
  {author} {\bibfnamefont {Q.}~\bibnamefont {Song}}, \bibinfo {author}
  {\bibfnamefont {Y.}~\bibnamefont {Liu}}, \bibinfo {author} {\bibfnamefont
  {Y.}~\bibnamefont {Wu}}, \bibinfo {author} {\bibfnamefont {S.}~\bibnamefont
  {Wang}}, \bibinfo {author} {\bibfnamefont {J.}~\bibnamefont {Yu}}, \bibinfo
  {author} {\bibfnamefont {J.}~\bibnamefont {Han}},\ and\ \bibinfo {author}
  {\bibfnamefont {D.-P.}\ \bibnamefont {Tsai}},\ }\bibfield  {title} {\bibinfo
  {title} {All-dielectric metasurface for high-performance structural color},\
  }\href@noop {} {\bibfield  {journal} {\bibinfo  {journal} {Nature
  communications}\ }\textbf {\bibinfo {volume} {11}},\ \bibinfo {pages} {1}
  (\bibinfo {year} {2020})}\BibitemShut {NoStop}%
\bibitem [{\citenamefont {Hemmatyar}\ \emph
  {et~al.}(2020{\natexlab{b}})\citenamefont {Hemmatyar}, \citenamefont {Lu},
  \citenamefont {Brown}, \citenamefont {Maleki},\ and\ \citenamefont
  {Adibi}}]{hemmatyar2020fano}%
  \BibitemOpen
  \bibfield  {author} {\bibinfo {author} {\bibfnamefont {O.}~\bibnamefont
  {Hemmatyar}}, \bibinfo {author} {\bibfnamefont {Z.}~\bibnamefont {Lu}},
  \bibinfo {author} {\bibfnamefont {T.}~\bibnamefont {Brown}}, \bibinfo
  {author} {\bibfnamefont {H.}~\bibnamefont {Maleki}},\ and\ \bibinfo {author}
  {\bibfnamefont {A.}~\bibnamefont {Adibi}},\ }\bibfield  {title} {\bibinfo
  {title} {Fano resonant all-dielectric metasurfaces for polarization-sensitive
  structural coloration},\ }in\ \href@noop {} {\emph {\bibinfo {booktitle}
  {2020 Conference on Lasers and Electro-Optics (CLEO)}}}\ (\bibinfo
  {organization} {IEEE},\ \bibinfo {year} {2020})\ pp.\ \bibinfo {pages}
  {1--2}\BibitemShut {NoStop}%
\bibitem [{\citenamefont {Franklin}\ \emph {et~al.}(2015)\citenamefont
  {Franklin}, \citenamefont {Chen}, \citenamefont {Vazquez-Guardado},
  \citenamefont {Modak}, \citenamefont {Boroumand}, \citenamefont {Xu},
  \citenamefont {Wu},\ and\ \citenamefont {Chanda}}]{franklin2015polarization}%
  \BibitemOpen
  \bibfield  {author} {\bibinfo {author} {\bibfnamefont {D.}~\bibnamefont
  {Franklin}}, \bibinfo {author} {\bibfnamefont {Y.}~\bibnamefont {Chen}},
  \bibinfo {author} {\bibfnamefont {A.}~\bibnamefont {Vazquez-Guardado}},
  \bibinfo {author} {\bibfnamefont {S.}~\bibnamefont {Modak}}, \bibinfo
  {author} {\bibfnamefont {J.}~\bibnamefont {Boroumand}}, \bibinfo {author}
  {\bibfnamefont {D.}~\bibnamefont {Xu}}, \bibinfo {author} {\bibfnamefont
  {S.-T.}\ \bibnamefont {Wu}},\ and\ \bibinfo {author} {\bibfnamefont
  {D.}~\bibnamefont {Chanda}},\ }\bibfield  {title} {\bibinfo {title}
  {Polarization-independent actively tunable colour generation on imprinted
  plasmonic surfaces},\ }\href@noop {} {\bibfield  {journal} {\bibinfo
  {journal} {Nature communications}\ }\textbf {\bibinfo {volume} {6}},\
  \bibinfo {pages} {1} (\bibinfo {year} {2015})}\BibitemShut {NoStop}%
\bibitem [{\citenamefont {Olson}\ \emph {et~al.}(2016)\citenamefont {Olson},
  \citenamefont {Manjavacas}, \citenamefont {Basu}, \citenamefont {Huang},
  \citenamefont {Schlather}, \citenamefont {Zheng}, \citenamefont {Halas},
  \citenamefont {Nordlander},\ and\ \citenamefont {Link}}]{olson2016high}%
  \BibitemOpen
  \bibfield  {author} {\bibinfo {author} {\bibfnamefont {J.}~\bibnamefont
  {Olson}}, \bibinfo {author} {\bibfnamefont {A.}~\bibnamefont {Manjavacas}},
  \bibinfo {author} {\bibfnamefont {T.}~\bibnamefont {Basu}}, \bibinfo {author}
  {\bibfnamefont {D.}~\bibnamefont {Huang}}, \bibinfo {author} {\bibfnamefont
  {A.~E.}\ \bibnamefont {Schlather}}, \bibinfo {author} {\bibfnamefont
  {B.}~\bibnamefont {Zheng}}, \bibinfo {author} {\bibfnamefont {N.~J.}\
  \bibnamefont {Halas}}, \bibinfo {author} {\bibfnamefont {P.}~\bibnamefont
  {Nordlander}},\ and\ \bibinfo {author} {\bibfnamefont {S.}~\bibnamefont
  {Link}},\ }\bibfield  {title} {\bibinfo {title} {High chromaticity aluminum
  plasmonic pixels for active liquid crystal displays},\ }\href@noop {}
  {\bibfield  {journal} {\bibinfo  {journal} {ACS nano}\ }\textbf {\bibinfo
  {volume} {10}},\ \bibinfo {pages} {1108} (\bibinfo {year}
  {2016})}\BibitemShut {NoStop}%
\bibitem [{\citenamefont {Tseng}\ \emph {et~al.}(2017)\citenamefont {Tseng},
  \citenamefont {Yang}, \citenamefont {Semmlinger}, \citenamefont {Zhang},
  \citenamefont {Nordlander},\ and\ \citenamefont {Halas}}]{tseng2017two}%
  \BibitemOpen
  \bibfield  {author} {\bibinfo {author} {\bibfnamefont {M.~L.}\ \bibnamefont
  {Tseng}}, \bibinfo {author} {\bibfnamefont {J.}~\bibnamefont {Yang}},
  \bibinfo {author} {\bibfnamefont {M.}~\bibnamefont {Semmlinger}}, \bibinfo
  {author} {\bibfnamefont {C.}~\bibnamefont {Zhang}}, \bibinfo {author}
  {\bibfnamefont {P.}~\bibnamefont {Nordlander}},\ and\ \bibinfo {author}
  {\bibfnamefont {N.~J.}\ \bibnamefont {Halas}},\ }\bibfield  {title} {\bibinfo
  {title} {Two-dimensional active tuning of an aluminum plasmonic array for
  full-spectrum response},\ }\href@noop {} {\bibfield  {journal} {\bibinfo
  {journal} {Nano letters}\ }\textbf {\bibinfo {volume} {17}},\ \bibinfo
  {pages} {6034} (\bibinfo {year} {2017})}\BibitemShut {NoStop}%
\bibitem [{\citenamefont {Song}\ \emph {et~al.}(2017)\citenamefont {Song},
  \citenamefont {Ma}, \citenamefont {Pu}, \citenamefont {Li}, \citenamefont
  {Liu}, \citenamefont {Gao}, \citenamefont {Zhao}, \citenamefont {Wang},
  \citenamefont {Wang},\ and\ \citenamefont {Luo}}]{song2017actively}%
  \BibitemOpen
  \bibfield  {author} {\bibinfo {author} {\bibfnamefont {S.}~\bibnamefont
  {Song}}, \bibinfo {author} {\bibfnamefont {X.}~\bibnamefont {Ma}}, \bibinfo
  {author} {\bibfnamefont {M.}~\bibnamefont {Pu}}, \bibinfo {author}
  {\bibfnamefont {X.}~\bibnamefont {Li}}, \bibinfo {author} {\bibfnamefont
  {K.}~\bibnamefont {Liu}}, \bibinfo {author} {\bibfnamefont {P.}~\bibnamefont
  {Gao}}, \bibinfo {author} {\bibfnamefont {Z.}~\bibnamefont {Zhao}}, \bibinfo
  {author} {\bibfnamefont {Y.}~\bibnamefont {Wang}}, \bibinfo {author}
  {\bibfnamefont {C.}~\bibnamefont {Wang}},\ and\ \bibinfo {author}
  {\bibfnamefont {X.}~\bibnamefont {Luo}},\ }\bibfield  {title} {\bibinfo
  {title} {Actively tunable structural color rendering with tensile
  substrate},\ }\href@noop {} {\bibfield  {journal} {\bibinfo  {journal}
  {Advanced Optical Materials}\ }\textbf {\bibinfo {volume} {5}},\ \bibinfo
  {pages} {1600829} (\bibinfo {year} {2017})}\BibitemShut {NoStop}%
\bibitem [{\citenamefont {Gutruf}\ \emph {et~al.}(2016)\citenamefont {Gutruf},
  \citenamefont {Zou}, \citenamefont {Withayachumnankul}, \citenamefont
  {Bhaskaran}, \citenamefont {Sriram},\ and\ \citenamefont
  {Fumeaux}}]{gutruf2016mechanically}%
  \BibitemOpen
  \bibfield  {author} {\bibinfo {author} {\bibfnamefont {P.}~\bibnamefont
  {Gutruf}}, \bibinfo {author} {\bibfnamefont {C.}~\bibnamefont {Zou}},
  \bibinfo {author} {\bibfnamefont {W.}~\bibnamefont {Withayachumnankul}},
  \bibinfo {author} {\bibfnamefont {M.}~\bibnamefont {Bhaskaran}}, \bibinfo
  {author} {\bibfnamefont {S.}~\bibnamefont {Sriram}},\ and\ \bibinfo {author}
  {\bibfnamefont {C.}~\bibnamefont {Fumeaux}},\ }\bibfield  {title} {\bibinfo
  {title} {Mechanically tunable dielectric resonator metasurfaces at visible
  frequencies},\ }\href@noop {} {\bibfield  {journal} {\bibinfo  {journal} {ACS
  nano}\ }\textbf {\bibinfo {volume} {10}},\ \bibinfo {pages} {133} (\bibinfo
  {year} {2016})}\BibitemShut {NoStop}%
\bibitem [{\citenamefont {Quan}\ \emph {et~al.}(2020)\citenamefont {Quan},
  \citenamefont {Kim}, \citenamefont {Kim}, \citenamefont {Min},\ and\
  \citenamefont {Ahn}}]{quan2020stretchable}%
  \BibitemOpen
  \bibfield  {author} {\bibinfo {author} {\bibfnamefont {Y.-J.}\ \bibnamefont
  {Quan}}, \bibinfo {author} {\bibfnamefont {Y.-G.}\ \bibnamefont {Kim}},
  \bibinfo {author} {\bibfnamefont {M.-S.}\ \bibnamefont {Kim}}, \bibinfo
  {author} {\bibfnamefont {S.-H.}\ \bibnamefont {Min}},\ and\ \bibinfo {author}
  {\bibfnamefont {S.-H.}\ \bibnamefont {Ahn}},\ }\bibfield  {title} {\bibinfo
  {title} {Stretchable biaxial and shear strain sensors using diffractive
  structural colors},\ }\href@noop {} {\bibfield  {journal} {\bibinfo
  {journal} {ACS nano}\ }\textbf {\bibinfo {volume} {14}},\ \bibinfo {pages}
  {5392} (\bibinfo {year} {2020})}\BibitemShut {NoStop}%
\bibitem [{\citenamefont {King}\ \emph {et~al.}(2015)\citenamefont {King},
  \citenamefont {Liu}, \citenamefont {Yang}, \citenamefont {Cerjan},
  \citenamefont {Everitt}, \citenamefont {Nordlander},\ and\ \citenamefont
  {Halas}}]{king2015fano}%
  \BibitemOpen
  \bibfield  {author} {\bibinfo {author} {\bibfnamefont {N.~S.}\ \bibnamefont
  {King}}, \bibinfo {author} {\bibfnamefont {L.}~\bibnamefont {Liu}}, \bibinfo
  {author} {\bibfnamefont {X.}~\bibnamefont {Yang}}, \bibinfo {author}
  {\bibfnamefont {B.}~\bibnamefont {Cerjan}}, \bibinfo {author} {\bibfnamefont
  {H.~O.}\ \bibnamefont {Everitt}}, \bibinfo {author} {\bibfnamefont
  {P.}~\bibnamefont {Nordlander}},\ and\ \bibinfo {author} {\bibfnamefont
  {N.~J.}\ \bibnamefont {Halas}},\ }\bibfield  {title} {\bibinfo {title} {Fano
  resonant aluminum nanoclusters for plasmonic colorimetric sensing},\
  }\href@noop {} {\bibfield  {journal} {\bibinfo  {journal} {ACS nano}\
  }\textbf {\bibinfo {volume} {9}},\ \bibinfo {pages} {10628} (\bibinfo {year}
  {2015})}\BibitemShut {NoStop}%
\bibitem [{\citenamefont {Chen}\ \emph {et~al.}(2017)\citenamefont {Chen},
  \citenamefont {Duan}, \citenamefont {Matuschek}, \citenamefont {Zhou},
  \citenamefont {Neubrech}, \citenamefont {Duan},\ and\ \citenamefont
  {Liu}}]{chen2017dynamic}%
  \BibitemOpen
  \bibfield  {author} {\bibinfo {author} {\bibfnamefont {Y.}~\bibnamefont
  {Chen}}, \bibinfo {author} {\bibfnamefont {X.}~\bibnamefont {Duan}}, \bibinfo
  {author} {\bibfnamefont {M.}~\bibnamefont {Matuschek}}, \bibinfo {author}
  {\bibfnamefont {Y.}~\bibnamefont {Zhou}}, \bibinfo {author} {\bibfnamefont
  {F.}~\bibnamefont {Neubrech}}, \bibinfo {author} {\bibfnamefont
  {H.}~\bibnamefont {Duan}},\ and\ \bibinfo {author} {\bibfnamefont
  {N.}~\bibnamefont {Liu}},\ }\bibfield  {title} {\bibinfo {title} {Dynamic
  color displays using stepwise cavity resonators},\ }\href@noop {} {\bibfield
  {journal} {\bibinfo  {journal} {Nano Letters}\ }\textbf {\bibinfo {volume}
  {17}},\ \bibinfo {pages} {5555} (\bibinfo {year} {2017})}\BibitemShut
  {NoStop}%
\bibitem [{\citenamefont {Yang}\ \emph {et~al.}(2018)\citenamefont {Yang},
  \citenamefont {Liu}, \citenamefont {Li}, \citenamefont {Cheng}, \citenamefont
  {Chen},\ and\ \citenamefont {Tian}}]{yang2018polarization}%
  \BibitemOpen
  \bibfield  {author} {\bibinfo {author} {\bibfnamefont {B.}~\bibnamefont
  {Yang}}, \bibinfo {author} {\bibfnamefont {W.}~\bibnamefont {Liu}}, \bibinfo
  {author} {\bibfnamefont {Z.}~\bibnamefont {Li}}, \bibinfo {author}
  {\bibfnamefont {H.}~\bibnamefont {Cheng}}, \bibinfo {author} {\bibfnamefont
  {S.}~\bibnamefont {Chen}},\ and\ \bibinfo {author} {\bibfnamefont
  {J.}~\bibnamefont {Tian}},\ }\bibfield  {title} {\bibinfo {title}
  {Polarization-sensitive structural colors with hue-and-saturation tuning
  based on all-dielectric nanopixels},\ }\href@noop {} {\bibfield  {journal}
  {\bibinfo  {journal} {Advanced Optical Materials}\ }\textbf {\bibinfo
  {volume} {6}},\ \bibinfo {pages} {1701009} (\bibinfo {year}
  {2018})}\BibitemShut {NoStop}%
\bibitem [{\citenamefont {Wuttig}\ \emph {et~al.}(2017)\citenamefont {Wuttig},
  \citenamefont {Bhaskaran},\ and\ \citenamefont {Taubner}}]{wuttig2017phase}%
  \BibitemOpen
  \bibfield  {author} {\bibinfo {author} {\bibfnamefont {M.}~\bibnamefont
  {Wuttig}}, \bibinfo {author} {\bibfnamefont {H.}~\bibnamefont {Bhaskaran}},\
  and\ \bibinfo {author} {\bibfnamefont {T.}~\bibnamefont {Taubner}},\
  }\bibfield  {title} {\bibinfo {title} {Phase-change materials for
  non-volatile photonic applications},\ }\href@noop {} {\bibfield  {journal}
  {\bibinfo  {journal} {Nature Photonics}\ }\textbf {\bibinfo {volume} {11}},\
  \bibinfo {pages} {465} (\bibinfo {year} {2017})}\BibitemShut {NoStop}%
\bibitem [{\citenamefont {Feldmann}\ \emph {et~al.}(2019)\citenamefont
  {Feldmann}, \citenamefont {Youngblood}, \citenamefont {Wright}, \citenamefont
  {Bhaskaran},\ and\ \citenamefont {Pernice}}]{feldmann2019all}%
  \BibitemOpen
  \bibfield  {author} {\bibinfo {author} {\bibfnamefont {J.}~\bibnamefont
  {Feldmann}}, \bibinfo {author} {\bibfnamefont {N.}~\bibnamefont
  {Youngblood}}, \bibinfo {author} {\bibfnamefont {C.~D.}\ \bibnamefont
  {Wright}}, \bibinfo {author} {\bibfnamefont {H.}~\bibnamefont {Bhaskaran}},\
  and\ \bibinfo {author} {\bibfnamefont {W.}~\bibnamefont {Pernice}},\
  }\bibfield  {title} {\bibinfo {title} {All-optical spiking neurosynaptic
  networks with self-learning capabilities},\ }\href@noop {} {\bibfield
  {journal} {\bibinfo  {journal} {Nature}\ }\textbf {\bibinfo {volume} {569}},\
  \bibinfo {pages} {208} (\bibinfo {year} {2019})}\BibitemShut {NoStop}%
\bibitem [{\citenamefont {Ding}\ \emph {et~al.}(2019)\citenamefont {Ding},
  \citenamefont {Yang},\ and\ \citenamefont {Bozhevolnyi}}]{ding2019dynamic}%
  \BibitemOpen
  \bibfield  {author} {\bibinfo {author} {\bibfnamefont {F.}~\bibnamefont
  {Ding}}, \bibinfo {author} {\bibfnamefont {Y.}~\bibnamefont {Yang}},\ and\
  \bibinfo {author} {\bibfnamefont {S.~I.}\ \bibnamefont {Bozhevolnyi}},\
  }\bibfield  {title} {\bibinfo {title} {Dynamic metasurfaces using
  phase-change chalcogenides},\ }\href@noop {} {\bibfield  {journal} {\bibinfo
  {journal} {Advanced Optical Materials}\ }\textbf {\bibinfo {volume} {7}},\
  \bibinfo {pages} {1801709} (\bibinfo {year} {2019})}\BibitemShut {NoStop}%
\bibitem [{\citenamefont {Abdollahramezani}\ \emph
  {et~al.}(2020{\natexlab{b}})\citenamefont {Abdollahramezani}, \citenamefont
  {Hemmatyar}, \citenamefont {Taghinejad}, \citenamefont {Krasnok},
  \citenamefont {Kiarashinejad}, \citenamefont {Zandehshahvar}, \citenamefont
  {Alù},\ and\ \citenamefont {Adibi}}]{abdollahramezani2020tunable}%
  \BibitemOpen
  \bibfield  {author} {\bibinfo {author} {\bibfnamefont {S.}~\bibnamefont
  {Abdollahramezani}}, \bibinfo {author} {\bibfnamefont {O.}~\bibnamefont
  {Hemmatyar}}, \bibinfo {author} {\bibfnamefont {H.}~\bibnamefont
  {Taghinejad}}, \bibinfo {author} {\bibfnamefont {A.}~\bibnamefont {Krasnok}},
  \bibinfo {author} {\bibfnamefont {Y.}~\bibnamefont {Kiarashinejad}}, \bibinfo
  {author} {\bibfnamefont {M.}~\bibnamefont {Zandehshahvar}}, \bibinfo {author}
  {\bibfnamefont {A.}~\bibnamefont {Alù}},\ and\ \bibinfo {author}
  {\bibfnamefont {A.}~\bibnamefont {Adibi}},\ }\bibfield  {title} {\bibinfo
  {title} {Tunable nanophotonics enabled by chalcogenide phase-change
  materials},\ }\href
  {https://doi.org/https://doi.org/10.1515/nanoph-2020-0039} {\bibfield
  {journal} {\bibinfo  {journal} {Nanophotonics}\ }\textbf {\bibinfo {volume}
  {9}},\ \bibinfo {pages} {1189 } (\bibinfo {year}
  {2020}{\natexlab{b}})}\BibitemShut {NoStop}%
\bibitem [{\citenamefont {Hail}\ \emph {et~al.}(2019)\citenamefont {Hail},
  \citenamefont {Michel}, \citenamefont {Poulikakos},\ and\ \citenamefont
  {Eghlidi}}]{hail2019optical}%
  \BibitemOpen
  \bibfield  {author} {\bibinfo {author} {\bibfnamefont {C.~U.}\ \bibnamefont
  {Hail}}, \bibinfo {author} {\bibfnamefont {A.-K.~U.}\ \bibnamefont {Michel}},
  \bibinfo {author} {\bibfnamefont {D.}~\bibnamefont {Poulikakos}},\ and\
  \bibinfo {author} {\bibfnamefont {H.}~\bibnamefont {Eghlidi}},\ }\bibfield
  {title} {\bibinfo {title} {Optical metasurfaces: evolving from passive to
  adaptive},\ }\href@noop {} {\bibfield  {journal} {\bibinfo  {journal}
  {Advanced Optical Materials}\ }\textbf {\bibinfo {volume} {7}},\ \bibinfo
  {pages} {1801786} (\bibinfo {year} {2019})}\BibitemShut {NoStop}%
\bibitem [{\citenamefont {Dong}\ \emph {et~al.}(2018)\citenamefont {Dong},
  \citenamefont {Qiu}, \citenamefont {Zhou}, \citenamefont {Banas},
  \citenamefont {Banas}, \citenamefont {Breese}, \citenamefont {Cao},\ and\
  \citenamefont {Simpson}}]{dong2018tunable}%
  \BibitemOpen
  \bibfield  {author} {\bibinfo {author} {\bibfnamefont {W.}~\bibnamefont
  {Dong}}, \bibinfo {author} {\bibfnamefont {Y.}~\bibnamefont {Qiu}}, \bibinfo
  {author} {\bibfnamefont {X.}~\bibnamefont {Zhou}}, \bibinfo {author}
  {\bibfnamefont {A.}~\bibnamefont {Banas}}, \bibinfo {author} {\bibfnamefont
  {K.}~\bibnamefont {Banas}}, \bibinfo {author} {\bibfnamefont {M.~B.}\
  \bibnamefont {Breese}}, \bibinfo {author} {\bibfnamefont {T.}~\bibnamefont
  {Cao}},\ and\ \bibinfo {author} {\bibfnamefont {R.~E.}\ \bibnamefont
  {Simpson}},\ }\bibfield  {title} {\bibinfo {title} {Tunable mid-infrared
  phase-change metasurface},\ }\href@noop {} {\bibfield  {journal} {\bibinfo
  {journal} {Advanced Optical Materials}\ }\textbf {\bibinfo {volume} {6}},\
  \bibinfo {pages} {1701346} (\bibinfo {year} {2018})}\BibitemShut {NoStop}%
\bibitem [{\citenamefont {Gholipour}\ \emph {et~al.}(2013)\citenamefont
  {Gholipour}, \citenamefont {Zhang}, \citenamefont {MacDonald}, \citenamefont
  {Hewak},\ and\ \citenamefont {Zheludev}}]{gholipour2013all}%
  \BibitemOpen
  \bibfield  {author} {\bibinfo {author} {\bibfnamefont {B.}~\bibnamefont
  {Gholipour}}, \bibinfo {author} {\bibfnamefont {J.}~\bibnamefont {Zhang}},
  \bibinfo {author} {\bibfnamefont {K.~F.}\ \bibnamefont {MacDonald}}, \bibinfo
  {author} {\bibfnamefont {D.~W.}\ \bibnamefont {Hewak}},\ and\ \bibinfo
  {author} {\bibfnamefont {N.~I.}\ \bibnamefont {Zheludev}},\ }\bibfield
  {title} {\bibinfo {title} {An all-optical, non-volatile, bidirectional,
  phase-change meta-switch},\ }\href@noop {} {\bibfield  {journal} {\bibinfo
  {journal} {Advanced materials}\ }\textbf {\bibinfo {volume} {25}},\ \bibinfo
  {pages} {3050} (\bibinfo {year} {2013})}\BibitemShut {NoStop}%
\bibitem [{\citenamefont {Michel}\ \emph {et~al.}(2013)\citenamefont {Michel},
  \citenamefont {Chigrin}, \citenamefont {Mass}, \citenamefont {Schonauer},
  \citenamefont {Salinga}, \citenamefont {Wuttig},\ and\ \citenamefont
  {Taubner}}]{michel2013using}%
  \BibitemOpen
  \bibfield  {author} {\bibinfo {author} {\bibfnamefont {A.-K.~U.}\
  \bibnamefont {Michel}}, \bibinfo {author} {\bibfnamefont {D.~N.}\
  \bibnamefont {Chigrin}}, \bibinfo {author} {\bibfnamefont {T.~W.}\
  \bibnamefont {Mass}}, \bibinfo {author} {\bibfnamefont {K.}~\bibnamefont
  {Schonauer}}, \bibinfo {author} {\bibfnamefont {M.}~\bibnamefont {Salinga}},
  \bibinfo {author} {\bibfnamefont {M.}~\bibnamefont {Wuttig}},\ and\ \bibinfo
  {author} {\bibfnamefont {T.}~\bibnamefont {Taubner}},\ }\bibfield  {title}
  {\bibinfo {title} {Using low-loss phase-change materials for mid-infrared
  antenna resonance tuning},\ }\href@noop {} {\bibfield  {journal} {\bibinfo
  {journal} {Nano letters}\ }\textbf {\bibinfo {volume} {13}},\ \bibinfo
  {pages} {3470} (\bibinfo {year} {2013})}\BibitemShut {NoStop}%
\bibitem [{\citenamefont {Abdollahramezani}\ \emph
  {et~al.}(2018{\natexlab{a}})\citenamefont {Abdollahramezani}, \citenamefont
  {Taghinejad}, \citenamefont {Fan}, \citenamefont {Kiarashinejad},
  \citenamefont {Eftekhar},\ and\ \citenamefont
  {Adibi}}]{abdollahramezani2018reconfigurable}%
  \BibitemOpen
  \bibfield  {author} {\bibinfo {author} {\bibfnamefont {S.}~\bibnamefont
  {Abdollahramezani}}, \bibinfo {author} {\bibfnamefont {H.}~\bibnamefont
  {Taghinejad}}, \bibinfo {author} {\bibfnamefont {T.}~\bibnamefont {Fan}},
  \bibinfo {author} {\bibfnamefont {Y.}~\bibnamefont {Kiarashinejad}}, \bibinfo
  {author} {\bibfnamefont {A.~A.}\ \bibnamefont {Eftekhar}},\ and\ \bibinfo
  {author} {\bibfnamefont {A.}~\bibnamefont {Adibi}},\ }\bibfield  {title}
  {\bibinfo {title} {Reconfigurable multifunctional metasurfaces employing
  hybrid phase-change plasmonic architecture},\ }\href@noop {} {\bibfield
  {journal} {\bibinfo  {journal} {arXiv preprint arXiv:1809.08907}\ } (\bibinfo
  {year} {2018}{\natexlab{a}})}\BibitemShut {NoStop}%
\bibitem [{\citenamefont {Zhang}\ \emph {et~al.}(2019)\citenamefont {Zhang},
  \citenamefont {Chou}, \citenamefont {Li}, \citenamefont {Li}, \citenamefont
  {Du}, \citenamefont {Yadav}, \citenamefont {Zhou}, \citenamefont
  {Shalaginov}, \citenamefont {Fang}, \citenamefont {Zhong} \emph
  {et~al.}}]{zhang2019broadband}%
  \BibitemOpen
  \bibfield  {author} {\bibinfo {author} {\bibfnamefont {Y.}~\bibnamefont
  {Zhang}}, \bibinfo {author} {\bibfnamefont {J.~B.}\ \bibnamefont {Chou}},
  \bibinfo {author} {\bibfnamefont {J.}~\bibnamefont {Li}}, \bibinfo {author}
  {\bibfnamefont {H.}~\bibnamefont {Li}}, \bibinfo {author} {\bibfnamefont
  {Q.}~\bibnamefont {Du}}, \bibinfo {author} {\bibfnamefont {A.}~\bibnamefont
  {Yadav}}, \bibinfo {author} {\bibfnamefont {S.}~\bibnamefont {Zhou}},
  \bibinfo {author} {\bibfnamefont {M.~Y.}\ \bibnamefont {Shalaginov}},
  \bibinfo {author} {\bibfnamefont {Z.}~\bibnamefont {Fang}}, \bibinfo {author}
  {\bibfnamefont {H.}~\bibnamefont {Zhong}}, \emph {et~al.},\ }\bibfield
  {title} {\bibinfo {title} {Broadband transparent optical phase change
  materials for high-performance nonvolatile photonics},\ }\href@noop {}
  {\bibfield  {journal} {\bibinfo  {journal} {Nature communications}\ }\textbf
  {\bibinfo {volume} {10}},\ \bibinfo {pages} {1} (\bibinfo {year}
  {2019})}\BibitemShut {NoStop}%
\bibitem [{\citenamefont {Taghinejad}\ \emph {et~al.}(2020)\citenamefont
  {Taghinejad}, \citenamefont {Abdollahramezani}, \citenamefont {Eftekhar},
  \citenamefont {Fan}, \citenamefont {Hosseinnia}, \citenamefont {Hemmatyar},
  \citenamefont {Dorche}, \citenamefont {Gallmon},\ and\ \citenamefont
  {Adibi}}]{taghinejad2020ito}%
  \BibitemOpen
  \bibfield  {author} {\bibinfo {author} {\bibfnamefont {H.}~\bibnamefont
  {Taghinejad}}, \bibinfo {author} {\bibfnamefont {S.}~\bibnamefont
  {Abdollahramezani}}, \bibinfo {author} {\bibfnamefont {A.~A.}\ \bibnamefont
  {Eftekhar}}, \bibinfo {author} {\bibfnamefont {T.}~\bibnamefont {Fan}},
  \bibinfo {author} {\bibfnamefont {A.~H.}\ \bibnamefont {Hosseinnia}},
  \bibinfo {author} {\bibfnamefont {O.}~\bibnamefont {Hemmatyar}}, \bibinfo
  {author} {\bibfnamefont {A.~E.}\ \bibnamefont {Dorche}}, \bibinfo {author}
  {\bibfnamefont {A.}~\bibnamefont {Gallmon}},\ and\ \bibinfo {author}
  {\bibfnamefont {A.}~\bibnamefont {Adibi}},\ }\bibfield  {title} {\bibinfo
  {title} {Ito-based mu-heaters for multi-stage switching of phase-change
  materials: Towards beyond-binary reconfigurable integrated photonics},\
  }\href@noop {} {\bibfield  {journal} {\bibinfo  {journal} {arXiv preprint
  arXiv:2003.04097}\ } (\bibinfo {year} {2020})}\BibitemShut {NoStop}%
\bibitem [{\citenamefont {Wang}\ \emph {et~al.}(2016)\citenamefont {Wang},
  \citenamefont {Rogers}, \citenamefont {Gholipour}, \citenamefont {Wang},
  \citenamefont {Yuan}, \citenamefont {Teng},\ and\ \citenamefont
  {Zheludev}}]{wang2016optically}%
  \BibitemOpen
  \bibfield  {author} {\bibinfo {author} {\bibfnamefont {Q.}~\bibnamefont
  {Wang}}, \bibinfo {author} {\bibfnamefont {E.~T.}\ \bibnamefont {Rogers}},
  \bibinfo {author} {\bibfnamefont {B.}~\bibnamefont {Gholipour}}, \bibinfo
  {author} {\bibfnamefont {C.-M.}\ \bibnamefont {Wang}}, \bibinfo {author}
  {\bibfnamefont {G.}~\bibnamefont {Yuan}}, \bibinfo {author} {\bibfnamefont
  {J.}~\bibnamefont {Teng}},\ and\ \bibinfo {author} {\bibfnamefont {N.~I.}\
  \bibnamefont {Zheludev}},\ }\bibfield  {title} {\bibinfo {title} {Optically
  reconfigurable metasurfaces and photonic devices based on phase change
  materials},\ }\href@noop {} {\bibfield  {journal} {\bibinfo  {journal}
  {Nature Photonics}\ }\textbf {\bibinfo {volume} {10}},\ \bibinfo {pages} {60}
  (\bibinfo {year} {2016})}\BibitemShut {NoStop}%
\bibitem [{\citenamefont {R{\'\i}os}\ \emph {et~al.}(2015)\citenamefont
  {R{\'\i}os}, \citenamefont {Stegmaier}, \citenamefont {Hosseini},
  \citenamefont {Wang}, \citenamefont {Scherer}, \citenamefont {Wright},
  \citenamefont {Bhaskaran},\ and\ \citenamefont
  {Pernice}}]{rios2015integrated}%
  \BibitemOpen
  \bibfield  {author} {\bibinfo {author} {\bibfnamefont {C.}~\bibnamefont
  {R{\'\i}os}}, \bibinfo {author} {\bibfnamefont {M.}~\bibnamefont
  {Stegmaier}}, \bibinfo {author} {\bibfnamefont {P.}~\bibnamefont {Hosseini}},
  \bibinfo {author} {\bibfnamefont {D.}~\bibnamefont {Wang}}, \bibinfo {author}
  {\bibfnamefont {T.}~\bibnamefont {Scherer}}, \bibinfo {author} {\bibfnamefont
  {C.~D.}\ \bibnamefont {Wright}}, \bibinfo {author} {\bibfnamefont
  {H.}~\bibnamefont {Bhaskaran}},\ and\ \bibinfo {author} {\bibfnamefont
  {W.~H.}\ \bibnamefont {Pernice}},\ }\bibfield  {title} {\bibinfo {title}
  {Integrated all-photonic non-volatile multi-level memory},\ }\href@noop {}
  {\bibfield  {journal} {\bibinfo  {journal} {Nature photonics}\ }\textbf
  {\bibinfo {volume} {9}},\ \bibinfo {pages} {725} (\bibinfo {year}
  {2015})}\BibitemShut {NoStop}%
\bibitem [{\citenamefont {Abdollahramezani}\ \emph
  {et~al.}(2021{\natexlab{a}})\citenamefont {Abdollahramezani}, \citenamefont
  {Hemmatyar}, \citenamefont {Taghinejad}, \citenamefont {Taghinejad},
  \citenamefont {Krasnok}, \citenamefont {Eftekhar}, \citenamefont {Teichrib},
  \citenamefont {Deshmukh}, \citenamefont {El-Sayed}, \citenamefont {Pop} \emph
  {et~al.}}]{abdollahramezani2021electrically}%
  \BibitemOpen
  \bibfield  {author} {\bibinfo {author} {\bibfnamefont {S.}~\bibnamefont
  {Abdollahramezani}}, \bibinfo {author} {\bibfnamefont {O.}~\bibnamefont
  {Hemmatyar}}, \bibinfo {author} {\bibfnamefont {M.}~\bibnamefont
  {Taghinejad}}, \bibinfo {author} {\bibfnamefont {H.}~\bibnamefont
  {Taghinejad}}, \bibinfo {author} {\bibfnamefont {A.}~\bibnamefont {Krasnok}},
  \bibinfo {author} {\bibfnamefont {A.~A.}\ \bibnamefont {Eftekhar}}, \bibinfo
  {author} {\bibfnamefont {C.}~\bibnamefont {Teichrib}}, \bibinfo {author}
  {\bibfnamefont {S.}~\bibnamefont {Deshmukh}}, \bibinfo {author}
  {\bibfnamefont {M.}~\bibnamefont {El-Sayed}}, \bibinfo {author}
  {\bibfnamefont {E.}~\bibnamefont {Pop}}, \emph {et~al.},\ }\bibfield  {title}
  {\bibinfo {title} {Electrically driven programmable phase-change meta-switch
  reaching 80\% efficiency},\ }\href@noop {} {\bibfield  {journal} {\bibinfo
  {journal} {arXiv preprint arXiv:2104.10381}\ } (\bibinfo {year}
  {2021}{\natexlab{a}})}\BibitemShut {NoStop}%
\bibitem [{\citenamefont {Krasnok}\ and\ \citenamefont
  {Al{\`u}}(2020)}]{krasnok2020active}%
  \BibitemOpen
  \bibfield  {author} {\bibinfo {author} {\bibfnamefont {A.}~\bibnamefont
  {Krasnok}}\ and\ \bibinfo {author} {\bibfnamefont {A.}~\bibnamefont
  {Al{\`u}}},\ }\bibfield  {title} {\bibinfo {title} {Active nanophotonics},\
  }\href@noop {} {\bibfield  {journal} {\bibinfo  {journal} {Proceedings of the
  IEEE}\ }\textbf {\bibinfo {volume} {108}},\ \bibinfo {pages} {628} (\bibinfo
  {year} {2020})}\BibitemShut {NoStop}%
\bibitem [{\citenamefont {Abdollahramezani}\ \emph
  {et~al.}(2018{\natexlab{b}})\citenamefont {Abdollahramezani}, \citenamefont
  {Taghinejad}, \citenamefont {Nejad}, \citenamefont {Eftekhar},\ and\
  \citenamefont {Adibi}}]{abdollahramezani2018dynamic}%
  \BibitemOpen
  \bibfield  {author} {\bibinfo {author} {\bibfnamefont {S.}~\bibnamefont
  {Abdollahramezani}}, \bibinfo {author} {\bibfnamefont {H.}~\bibnamefont
  {Taghinejad}}, \bibinfo {author} {\bibfnamefont {Y.~K.}\ \bibnamefont
  {Nejad}}, \bibinfo {author} {\bibfnamefont {A.~A.}\ \bibnamefont
  {Eftekhar}},\ and\ \bibinfo {author} {\bibfnamefont {A.}~\bibnamefont
  {Adibi}},\ }\bibfield  {title} {\bibinfo {title} {Dynamic dielectric
  metasurfaces incorporating phase-change material},\ }in\ \href@noop {} {\emph
  {\bibinfo {booktitle} {CLEO: Science and Innovations}}}\ (\bibinfo
  {organization} {Optical Society of America},\ \bibinfo {year} {2018})\ pp.\
  \bibinfo {pages} {SF1J--1}\BibitemShut {NoStop}%
\bibitem [{\citenamefont {Zheng}\ \emph {et~al.}(2018)\citenamefont {Zheng},
  \citenamefont {Khanolkar}, \citenamefont {Xu}, \citenamefont {Colburn},
  \citenamefont {Deshmukh}, \citenamefont {Myers}, \citenamefont {Frantz},
  \citenamefont {Pop}, \citenamefont {Hendrickson}, \citenamefont {Doylend}
  \emph {et~al.}}]{zheng2018gst}%
  \BibitemOpen
  \bibfield  {author} {\bibinfo {author} {\bibfnamefont {J.}~\bibnamefont
  {Zheng}}, \bibinfo {author} {\bibfnamefont {A.}~\bibnamefont {Khanolkar}},
  \bibinfo {author} {\bibfnamefont {P.}~\bibnamefont {Xu}}, \bibinfo {author}
  {\bibfnamefont {S.}~\bibnamefont {Colburn}}, \bibinfo {author} {\bibfnamefont
  {S.}~\bibnamefont {Deshmukh}}, \bibinfo {author} {\bibfnamefont
  {J.}~\bibnamefont {Myers}}, \bibinfo {author} {\bibfnamefont
  {J.}~\bibnamefont {Frantz}}, \bibinfo {author} {\bibfnamefont
  {E.}~\bibnamefont {Pop}}, \bibinfo {author} {\bibfnamefont {J.}~\bibnamefont
  {Hendrickson}}, \bibinfo {author} {\bibfnamefont {J.}~\bibnamefont
  {Doylend}}, \emph {et~al.},\ }\bibfield  {title} {\bibinfo {title}
  {Gst-on-silicon hybrid nanophotonic integrated circuits: a non-volatile
  quasi-continuously reprogrammable platform},\ }\href@noop {} {\bibfield
  {journal} {\bibinfo  {journal} {Optical Materials Express}\ }\textbf
  {\bibinfo {volume} {8}},\ \bibinfo {pages} {1551} (\bibinfo {year}
  {2018})}\BibitemShut {NoStop}%
\bibitem [{\citenamefont {Tian}\ \emph {et~al.}(2019)\citenamefont {Tian},
  \citenamefont {Luo}, \citenamefont {Yang}, \citenamefont {Ding},
  \citenamefont {Qu}, \citenamefont {Zhao}, \citenamefont {Qiu},\ and\
  \citenamefont {Bozhevolnyi}}]{tian2019active}%
  \BibitemOpen
  \bibfield  {author} {\bibinfo {author} {\bibfnamefont {J.}~\bibnamefont
  {Tian}}, \bibinfo {author} {\bibfnamefont {H.}~\bibnamefont {Luo}}, \bibinfo
  {author} {\bibfnamefont {Y.}~\bibnamefont {Yang}}, \bibinfo {author}
  {\bibfnamefont {F.}~\bibnamefont {Ding}}, \bibinfo {author} {\bibfnamefont
  {Y.}~\bibnamefont {Qu}}, \bibinfo {author} {\bibfnamefont {D.}~\bibnamefont
  {Zhao}}, \bibinfo {author} {\bibfnamefont {M.}~\bibnamefont {Qiu}},\ and\
  \bibinfo {author} {\bibfnamefont {S.~I.}\ \bibnamefont {Bozhevolnyi}},\
  }\bibfield  {title} {\bibinfo {title} {Active control of anapole states by
  structuring the phase-change alloy ge 2 sb 2 te 5},\ }\href@noop {}
  {\bibfield  {journal} {\bibinfo  {journal} {Nature communications}\ }\textbf
  {\bibinfo {volume} {10}},\ \bibinfo {pages} {1} (\bibinfo {year}
  {2019})}\BibitemShut {NoStop}%
\bibitem [{\citenamefont {Michel}\ \emph {et~al.}(2019)\citenamefont {Michel},
  \citenamefont {He{\ss}ler}, \citenamefont {Meyer}, \citenamefont {Pries},
  \citenamefont {Yu}, \citenamefont {Kalix}, \citenamefont {Lewin},
  \citenamefont {Hanss}, \citenamefont {De~Rose}, \citenamefont {Ma{\ss}} \emph
  {et~al.}}]{michel2019advanced}%
  \BibitemOpen
  \bibfield  {author} {\bibinfo {author} {\bibfnamefont {A.-K.~U.}\
  \bibnamefont {Michel}}, \bibinfo {author} {\bibfnamefont {A.}~\bibnamefont
  {He{\ss}ler}}, \bibinfo {author} {\bibfnamefont {S.}~\bibnamefont {Meyer}},
  \bibinfo {author} {\bibfnamefont {J.}~\bibnamefont {Pries}}, \bibinfo
  {author} {\bibfnamefont {Y.}~\bibnamefont {Yu}}, \bibinfo {author}
  {\bibfnamefont {T.}~\bibnamefont {Kalix}}, \bibinfo {author} {\bibfnamefont
  {M.}~\bibnamefont {Lewin}}, \bibinfo {author} {\bibfnamefont
  {J.}~\bibnamefont {Hanss}}, \bibinfo {author} {\bibfnamefont
  {A.}~\bibnamefont {De~Rose}}, \bibinfo {author} {\bibfnamefont {T.~W.}\
  \bibnamefont {Ma{\ss}}}, \emph {et~al.},\ }\bibfield  {title} {\bibinfo
  {title} {Advanced optical programming of individual meta-atoms beyond the
  effective medium approach},\ }\href@noop {} {\bibfield  {journal} {\bibinfo
  {journal} {Advanced Materials}\ }\textbf {\bibinfo {volume} {31}},\ \bibinfo
  {pages} {1901033} (\bibinfo {year} {2019})}\BibitemShut {NoStop}%
\bibitem [{\citenamefont {Xu}\ \emph {et~al.}(2019)\citenamefont {Xu},
  \citenamefont {Zheng}, \citenamefont {Doylend},\ and\ \citenamefont
  {Majumdar}}]{xu2019low}%
  \BibitemOpen
  \bibfield  {author} {\bibinfo {author} {\bibfnamefont {P.}~\bibnamefont
  {Xu}}, \bibinfo {author} {\bibfnamefont {J.}~\bibnamefont {Zheng}}, \bibinfo
  {author} {\bibfnamefont {J.~K.}\ \bibnamefont {Doylend}},\ and\ \bibinfo
  {author} {\bibfnamefont {A.}~\bibnamefont {Majumdar}},\ }\bibfield  {title}
  {\bibinfo {title} {Low-loss and broadband nonvolatile phase-change
  directional coupler switches},\ }\href@noop {} {\bibfield  {journal}
  {\bibinfo  {journal} {ACS Photonics}\ }\textbf {\bibinfo {volume} {6}},\
  \bibinfo {pages} {553} (\bibinfo {year} {2019})}\BibitemShut {NoStop}%
\bibitem [{\citenamefont {Abdollahramezani}\ \emph
  {et~al.}(2021{\natexlab{b}})\citenamefont {Abdollahramezani}, \citenamefont
  {Hemmatyar}, \citenamefont {Taghinejad}, \citenamefont {Taghinejad},
  \citenamefont {Kiarashinejad}, \citenamefont {Zandehshahvar}, \citenamefont
  {Fan}, \citenamefont {Deshmukh}, \citenamefont {Eftekhar}, \citenamefont
  {Cai} \emph {et~al.}}]{abdollahramezani2021dynamic}%
  \BibitemOpen
  \bibfield  {author} {\bibinfo {author} {\bibfnamefont {S.}~\bibnamefont
  {Abdollahramezani}}, \bibinfo {author} {\bibfnamefont {O.}~\bibnamefont
  {Hemmatyar}}, \bibinfo {author} {\bibfnamefont {M.}~\bibnamefont
  {Taghinejad}}, \bibinfo {author} {\bibfnamefont {H.}~\bibnamefont
  {Taghinejad}}, \bibinfo {author} {\bibfnamefont {Y.}~\bibnamefont
  {Kiarashinejad}}, \bibinfo {author} {\bibfnamefont {M.}~\bibnamefont
  {Zandehshahvar}}, \bibinfo {author} {\bibfnamefont {T.}~\bibnamefont {Fan}},
  \bibinfo {author} {\bibfnamefont {S.}~\bibnamefont {Deshmukh}}, \bibinfo
  {author} {\bibfnamefont {A.~A.}\ \bibnamefont {Eftekhar}}, \bibinfo {author}
  {\bibfnamefont {W.}~\bibnamefont {Cai}}, \emph {et~al.},\ }\bibfield  {title}
  {\bibinfo {title} {Dynamic hybrid metasurfaces},\ }\href@noop {} {\bibfield
  {journal} {\bibinfo  {journal} {Nano Letters}\ }\textbf {\bibinfo {volume}
  {21}},\ \bibinfo {pages} {1238} (\bibinfo {year}
  {2021}{\natexlab{b}})}\BibitemShut {NoStop}%
\bibitem [{\citenamefont {Wu}\ \emph {et~al.}(2021)\citenamefont {Wu},
  \citenamefont {Yu}, \citenamefont {Lee}, \citenamefont {Peng}, \citenamefont
  {Takeuchi},\ and\ \citenamefont {Li}}]{wu2021programmable}%
  \BibitemOpen
  \bibfield  {author} {\bibinfo {author} {\bibfnamefont {C.}~\bibnamefont
  {Wu}}, \bibinfo {author} {\bibfnamefont {H.}~\bibnamefont {Yu}}, \bibinfo
  {author} {\bibfnamefont {S.}~\bibnamefont {Lee}}, \bibinfo {author}
  {\bibfnamefont {R.}~\bibnamefont {Peng}}, \bibinfo {author} {\bibfnamefont
  {I.}~\bibnamefont {Takeuchi}},\ and\ \bibinfo {author} {\bibfnamefont
  {M.}~\bibnamefont {Li}},\ }\bibfield  {title} {\bibinfo {title} {Programmable
  phase-change metasurfaces on waveguides for multimode photonic convolutional
  neural network},\ }\href@noop {} {\bibfield  {journal} {\bibinfo  {journal}
  {Nature Communications}\ }\textbf {\bibinfo {volume} {12}},\ \bibinfo {pages}
  {1} (\bibinfo {year} {2021})}\BibitemShut {NoStop}%
\bibitem [{\citenamefont {Hemmatyar}\ \emph
  {et~al.}(2020{\natexlab{c}})\citenamefont {Hemmatyar}, \citenamefont
  {Abdollahramezani}, \citenamefont {Taghinejad},\ and\ \citenamefont
  {Adibi}}]{hemmatyar2020mixed}%
  \BibitemOpen
  \bibfield  {author} {\bibinfo {author} {\bibfnamefont {O.}~\bibnamefont
  {Hemmatyar}}, \bibinfo {author} {\bibfnamefont {S.}~\bibnamefont
  {Abdollahramezani}}, \bibinfo {author} {\bibfnamefont {H.}~\bibnamefont
  {Taghinejad}},\ and\ \bibinfo {author} {\bibfnamefont {A.}~\bibnamefont
  {Adibi}},\ }\bibfield  {title} {\bibinfo {title} {Mixed eletro-optic
  metasurface with a hybrid plasmonic-phase-change material architecture},\
  }in\ \href@noop {} {\emph {\bibinfo {booktitle} {CLEO: QELS\_Fundamental
  Science}}}\ (\bibinfo {organization} {Optical Society of America},\ \bibinfo
  {year} {2020})\ pp.\ \bibinfo {pages} {FW3Q--2}\BibitemShut {NoStop}%
\bibitem [{\citenamefont {de~Galarreta}\ \emph {et~al.}(2018)\citenamefont
  {de~Galarreta}, \citenamefont {Alexeev}, \citenamefont {Au}, \citenamefont
  {Lopez-Garcia}, \citenamefont {Klemm}, \citenamefont {Cryan}, \citenamefont
  {Bertolotti},\ and\ \citenamefont {Wright}}]{de2018nonvolatile}%
  \BibitemOpen
  \bibfield  {author} {\bibinfo {author} {\bibfnamefont {C.~R.}\ \bibnamefont
  {de~Galarreta}}, \bibinfo {author} {\bibfnamefont {A.~M.}\ \bibnamefont
  {Alexeev}}, \bibinfo {author} {\bibfnamefont {Y.-Y.}\ \bibnamefont {Au}},
  \bibinfo {author} {\bibfnamefont {M.}~\bibnamefont {Lopez-Garcia}}, \bibinfo
  {author} {\bibfnamefont {M.}~\bibnamefont {Klemm}}, \bibinfo {author}
  {\bibfnamefont {M.}~\bibnamefont {Cryan}}, \bibinfo {author} {\bibfnamefont
  {J.}~\bibnamefont {Bertolotti}},\ and\ \bibinfo {author} {\bibfnamefont
  {C.~D.}\ \bibnamefont {Wright}},\ }\bibfield  {title} {\bibinfo {title}
  {Nonvolatile reconfigurable phase-change metadevices for beam steering in the
  near infrared},\ }\href@noop {} {\bibfield  {journal} {\bibinfo  {journal}
  {Advanced Functional Materials}\ }\textbf {\bibinfo {volume} {28}},\ \bibinfo
  {pages} {1704993} (\bibinfo {year} {2018})}\BibitemShut {NoStop}%
\bibitem [{\citenamefont {Feldmann}\ \emph {et~al.}(2021)\citenamefont
  {Feldmann}, \citenamefont {Youngblood}, \citenamefont {Karpov}, \citenamefont
  {Gehring}, \citenamefont {Li}, \citenamefont {Stappers}, \citenamefont
  {Le~Gallo}, \citenamefont {Fu}, \citenamefont {Lukashchuk}, \citenamefont
  {Raja} \emph {et~al.}}]{feldmann2021parallel}%
  \BibitemOpen
  \bibfield  {author} {\bibinfo {author} {\bibfnamefont {J.}~\bibnamefont
  {Feldmann}}, \bibinfo {author} {\bibfnamefont {N.}~\bibnamefont
  {Youngblood}}, \bibinfo {author} {\bibfnamefont {M.}~\bibnamefont {Karpov}},
  \bibinfo {author} {\bibfnamefont {H.}~\bibnamefont {Gehring}}, \bibinfo
  {author} {\bibfnamefont {X.}~\bibnamefont {Li}}, \bibinfo {author}
  {\bibfnamefont {M.}~\bibnamefont {Stappers}}, \bibinfo {author}
  {\bibfnamefont {M.}~\bibnamefont {Le~Gallo}}, \bibinfo {author}
  {\bibfnamefont {X.}~\bibnamefont {Fu}}, \bibinfo {author} {\bibfnamefont
  {A.}~\bibnamefont {Lukashchuk}}, \bibinfo {author} {\bibfnamefont {A.~S.}\
  \bibnamefont {Raja}}, \emph {et~al.},\ }\bibfield  {title} {\bibinfo {title}
  {Parallel convolutional processing using an integrated photonic tensor
  core},\ }\href@noop {} {\bibfield  {journal} {\bibinfo  {journal} {Nature}\
  }\textbf {\bibinfo {volume} {589}},\ \bibinfo {pages} {52} (\bibinfo {year}
  {2021})}\BibitemShut {NoStop}%
\bibitem [{\citenamefont {Abdollahramezani}\ \emph
  {et~al.}(2020{\natexlab{c}})\citenamefont {Abdollahramezani}, \citenamefont
  {Kiarashinejad}, \citenamefont {Hemmatyar}, \citenamefont {Zandehshavar},\
  and\ \citenamefont {Adibi}}]{abdollahramezani2020electrically}%
  \BibitemOpen
  \bibfield  {author} {\bibinfo {author} {\bibfnamefont {S.}~\bibnamefont
  {Abdollahramezani}}, \bibinfo {author} {\bibfnamefont {Y.}~\bibnamefont
  {Kiarashinejad}}, \bibinfo {author} {\bibfnamefont {O.}~\bibnamefont
  {Hemmatyar}}, \bibinfo {author} {\bibfnamefont {M.}~\bibnamefont
  {Zandehshavar}},\ and\ \bibinfo {author} {\bibfnamefont {A.}~\bibnamefont
  {Adibi}},\ }\bibfield  {title} {\bibinfo {title} {Electrically programmable
  phased-array antenna using phase-change materials},\ }in\ \href@noop {}
  {\emph {\bibinfo {booktitle} {CLEO: QELS\_Fundamental Science}}}\ (\bibinfo
  {organization} {Optical Society of America},\ \bibinfo {year} {2020})\ pp.\
  \bibinfo {pages} {FW3Q--5}\BibitemShut {NoStop}%
\bibitem [{\citenamefont {Zheng}\ \emph {et~al.}(2020)\citenamefont {Zheng},
  \citenamefont {Fang}, \citenamefont {Wu}, \citenamefont {Zhu}, \citenamefont
  {Xu}, \citenamefont {Doylend}, \citenamefont {Deshmukh}, \citenamefont {Pop},
  \citenamefont {Dunham}, \citenamefont {Li} \emph
  {et~al.}}]{zheng2020nonvolatile}%
  \BibitemOpen
  \bibfield  {author} {\bibinfo {author} {\bibfnamefont {J.}~\bibnamefont
  {Zheng}}, \bibinfo {author} {\bibfnamefont {Z.}~\bibnamefont {Fang}},
  \bibinfo {author} {\bibfnamefont {C.}~\bibnamefont {Wu}}, \bibinfo {author}
  {\bibfnamefont {S.}~\bibnamefont {Zhu}}, \bibinfo {author} {\bibfnamefont
  {P.}~\bibnamefont {Xu}}, \bibinfo {author} {\bibfnamefont {J.~K.}\
  \bibnamefont {Doylend}}, \bibinfo {author} {\bibfnamefont {S.}~\bibnamefont
  {Deshmukh}}, \bibinfo {author} {\bibfnamefont {E.}~\bibnamefont {Pop}},
  \bibinfo {author} {\bibfnamefont {S.}~\bibnamefont {Dunham}}, \bibinfo
  {author} {\bibfnamefont {M.}~\bibnamefont {Li}}, \emph {et~al.},\ }\bibfield
  {title} {\bibinfo {title} {Nonvolatile electrically reconfigurable integrated
  photonic switch enabled by a silicon pin diode heater},\ }\href@noop {}
  {\bibfield  {journal} {\bibinfo  {journal} {Advanced Materials}\ }\textbf
  {\bibinfo {volume} {32}},\ \bibinfo {pages} {2001218} (\bibinfo {year}
  {2020})}\BibitemShut {NoStop}%
\bibitem [{\citenamefont {Leitis}\ \emph {et~al.}(2020)\citenamefont {Leitis},
  \citenamefont {He{\ss}ler}, \citenamefont {Wahl}, \citenamefont {Wuttig},
  \citenamefont {Taubner}, \citenamefont {Tittl},\ and\ \citenamefont
  {Altug}}]{leitis2020all}%
  \BibitemOpen
  \bibfield  {author} {\bibinfo {author} {\bibfnamefont {A.}~\bibnamefont
  {Leitis}}, \bibinfo {author} {\bibfnamefont {A.}~\bibnamefont {He{\ss}ler}},
  \bibinfo {author} {\bibfnamefont {S.}~\bibnamefont {Wahl}}, \bibinfo {author}
  {\bibfnamefont {M.}~\bibnamefont {Wuttig}}, \bibinfo {author} {\bibfnamefont
  {T.}~\bibnamefont {Taubner}}, \bibinfo {author} {\bibfnamefont
  {A.}~\bibnamefont {Tittl}},\ and\ \bibinfo {author} {\bibfnamefont
  {H.}~\bibnamefont {Altug}},\ }\bibfield  {title} {\bibinfo {title}
  {All-dielectric programmable huygens' metasurfaces},\ }\href@noop {}
  {\bibfield  {journal} {\bibinfo  {journal} {Advanced Functional Materials}\
  }\textbf {\bibinfo {volume} {30}},\ \bibinfo {pages} {1910259} (\bibinfo
  {year} {2020})}\BibitemShut {NoStop}%
\bibitem [{\citenamefont {Hemmatyar}\ \emph
  {et~al.}(2020{\natexlab{d}})\citenamefont {Hemmatyar}, \citenamefont
  {Abdollahramezani}, \citenamefont {Taghinejad},\ and\ \citenamefont
  {Adibi}}]{hemmatyar2020tunable2}%
  \BibitemOpen
  \bibfield  {author} {\bibinfo {author} {\bibfnamefont {O.}~\bibnamefont
  {Hemmatyar}}, \bibinfo {author} {\bibfnamefont {S.}~\bibnamefont
  {Abdollahramezani}}, \bibinfo {author} {\bibfnamefont {H.}~\bibnamefont
  {Taghinejad}},\ and\ \bibinfo {author} {\bibfnamefont {A.}~\bibnamefont
  {Adibi}},\ }\bibfield  {title} {\bibinfo {title} {Tunable
  polarization-independent absorber using a hybrid plasmonic and phase-change
  chalcogenide platform},\ }in\ \href@noop {} {\emph {\bibinfo {booktitle}
  {CLEO: QELS\_Fundamental Science}}}\ (\bibinfo {organization} {Optical
  Society of America},\ \bibinfo {year} {2020})\ pp.\ \bibinfo {pages}
  {FM3B--8}\BibitemShut {NoStop}%
\bibitem [{\citenamefont {Zhu}\ \emph {et~al.}(2020)\citenamefont {Zhu},
  \citenamefont {Abdollahramezani}, \citenamefont {Hemmatyar},\ and\
  \citenamefont {Adibi}}]{zhu2020linear}%
  \BibitemOpen
  \bibfield  {author} {\bibinfo {author} {\bibfnamefont {M.}~\bibnamefont
  {Zhu}}, \bibinfo {author} {\bibfnamefont {S.}~\bibnamefont
  {Abdollahramezani}}, \bibinfo {author} {\bibfnamefont {O.}~\bibnamefont
  {Hemmatyar}},\ and\ \bibinfo {author} {\bibfnamefont {A.}~\bibnamefont
  {Adibi}},\ }\bibfield  {title} {\bibinfo {title} {Linear and nonlinear
  focusing using reconfigurable all-dielectric metalens based on phase-change
  materials},\ }in\ \href@noop {} {\emph {\bibinfo {booktitle} {Laser
  Science}}}\ (\bibinfo {organization} {Optical Society of America},\ \bibinfo
  {year} {2020})\ pp.\ \bibinfo {pages} {JW6B--6}\BibitemShut {NoStop}%
\bibitem [{\citenamefont {Zhu}\ \emph {et~al.}(2021)\citenamefont {Zhu},
  \citenamefont {Abdollahramezani}, \citenamefont {Hemmatyar},\ and\
  \citenamefont {Adibi}}]{zhu2021tunable}%
  \BibitemOpen
  \bibfield  {author} {\bibinfo {author} {\bibfnamefont {M.}~\bibnamefont
  {Zhu}}, \bibinfo {author} {\bibfnamefont {S.}~\bibnamefont
  {Abdollahramezani}}, \bibinfo {author} {\bibfnamefont {O.}~\bibnamefont
  {Hemmatyar}},\ and\ \bibinfo {author} {\bibfnamefont {A.}~\bibnamefont
  {Adibi}},\ }\bibfield  {title} {\bibinfo {title} {Tunable third-harmonic
  generation using low-loss phase change chalcogenides},\ }in\ \href@noop {}
  {\emph {\bibinfo {booktitle} {Photonic and Phononic Properties of Engineered
  Nanostructures XI}}},\ Vol.\ \bibinfo {volume} {11694}\ (\bibinfo
  {organization} {International Society for Optics and Photonics},\ \bibinfo
  {year} {2021})\ p.\ \bibinfo {pages} {116941V}\BibitemShut {NoStop}%
\bibitem [{\citenamefont {Shalaginov}\ \emph {et~al.}(2021)\citenamefont
  {Shalaginov}, \citenamefont {An}, \citenamefont {Zhang}, \citenamefont
  {Yang}, \citenamefont {Su}, \citenamefont {Liberman}, \citenamefont {Chou},
  \citenamefont {Roberts}, \citenamefont {Kang}, \citenamefont {Rios} \emph
  {et~al.}}]{shalaginov2021reconfigurable}%
  \BibitemOpen
  \bibfield  {author} {\bibinfo {author} {\bibfnamefont {M.~Y.}\ \bibnamefont
  {Shalaginov}}, \bibinfo {author} {\bibfnamefont {S.}~\bibnamefont {An}},
  \bibinfo {author} {\bibfnamefont {Y.}~\bibnamefont {Zhang}}, \bibinfo
  {author} {\bibfnamefont {F.}~\bibnamefont {Yang}}, \bibinfo {author}
  {\bibfnamefont {P.}~\bibnamefont {Su}}, \bibinfo {author} {\bibfnamefont
  {V.}~\bibnamefont {Liberman}}, \bibinfo {author} {\bibfnamefont {J.~B.}\
  \bibnamefont {Chou}}, \bibinfo {author} {\bibfnamefont {C.~M.}\ \bibnamefont
  {Roberts}}, \bibinfo {author} {\bibfnamefont {M.}~\bibnamefont {Kang}},
  \bibinfo {author} {\bibfnamefont {C.}~\bibnamefont {Rios}}, \emph {et~al.},\
  }\bibfield  {title} {\bibinfo {title} {Reconfigurable all-dielectric metalens
  with diffraction-limited performance},\ }\href@noop {} {\bibfield  {journal}
  {\bibinfo  {journal} {Nature communications}\ }\textbf {\bibinfo {volume}
  {12}},\ \bibinfo {pages} {1} (\bibinfo {year} {2021})}\BibitemShut {NoStop}%
\bibitem [{\citenamefont {Zhang}\ \emph {et~al.}(2021)\citenamefont {Zhang},
  \citenamefont {Fowler}, \citenamefont {Liang}, \citenamefont {Azhar},
  \citenamefont {Shalaginov}, \citenamefont {Deckoff-Jones}, \citenamefont
  {An}, \citenamefont {Chou}, \citenamefont {Roberts}, \citenamefont {Liberman}
  \emph {et~al.}}]{zhang2021electrically}%
  \BibitemOpen
  \bibfield  {author} {\bibinfo {author} {\bibfnamefont {Y.}~\bibnamefont
  {Zhang}}, \bibinfo {author} {\bibfnamefont {C.}~\bibnamefont {Fowler}},
  \bibinfo {author} {\bibfnamefont {J.}~\bibnamefont {Liang}}, \bibinfo
  {author} {\bibfnamefont {B.}~\bibnamefont {Azhar}}, \bibinfo {author}
  {\bibfnamefont {M.~Y.}\ \bibnamefont {Shalaginov}}, \bibinfo {author}
  {\bibfnamefont {S.}~\bibnamefont {Deckoff-Jones}}, \bibinfo {author}
  {\bibfnamefont {S.}~\bibnamefont {An}}, \bibinfo {author} {\bibfnamefont
  {J.~B.}\ \bibnamefont {Chou}}, \bibinfo {author} {\bibfnamefont {C.~M.}\
  \bibnamefont {Roberts}}, \bibinfo {author} {\bibfnamefont {V.}~\bibnamefont
  {Liberman}}, \emph {et~al.},\ }\bibfield  {title} {\bibinfo {title}
  {Electrically reconfigurable non-volatile metasurface using low-loss optical
  phase-change material},\ }\href@noop {} {\bibfield  {journal} {\bibinfo
  {journal} {Nature Nanotechnology}\ ,\ \bibinfo {pages} {1}} (\bibinfo {year}
  {2021})}\BibitemShut {NoStop}%
\bibitem [{\citenamefont {Wang}\ \emph {et~al.}(2021)\citenamefont {Wang},
  \citenamefont {Landreman}, \citenamefont {Schoen}, \citenamefont {Okabe},
  \citenamefont {Marshall}, \citenamefont {Celano}, \citenamefont {Wong},
  \citenamefont {Park},\ and\ \citenamefont {Brongersma}}]{wang2021electrical}%
  \BibitemOpen
  \bibfield  {author} {\bibinfo {author} {\bibfnamefont {Y.}~\bibnamefont
  {Wang}}, \bibinfo {author} {\bibfnamefont {P.}~\bibnamefont {Landreman}},
  \bibinfo {author} {\bibfnamefont {D.}~\bibnamefont {Schoen}}, \bibinfo
  {author} {\bibfnamefont {K.}~\bibnamefont {Okabe}}, \bibinfo {author}
  {\bibfnamefont {A.}~\bibnamefont {Marshall}}, \bibinfo {author}
  {\bibfnamefont {U.}~\bibnamefont {Celano}}, \bibinfo {author} {\bibfnamefont
  {H.-S.~P.}\ \bibnamefont {Wong}}, \bibinfo {author} {\bibfnamefont
  {J.}~\bibnamefont {Park}},\ and\ \bibinfo {author} {\bibfnamefont {M.~L.}\
  \bibnamefont {Brongersma}},\ }\bibfield  {title} {\bibinfo {title}
  {Electrical tuning of phase-change antennas and metasurfaces},\ }\href@noop
  {} {\bibfield  {journal} {\bibinfo  {journal} {Nature Nanotechnology}\ ,\
  \bibinfo {pages} {1}} (\bibinfo {year} {2021})}\BibitemShut {NoStop}%
\bibitem [{\citenamefont {Lepeshov}\ and\ \citenamefont
  {Krasnok}(2021)}]{lepeshov2021tunable}%
  \BibitemOpen
  \bibfield  {author} {\bibinfo {author} {\bibfnamefont {S.}~\bibnamefont
  {Lepeshov}}\ and\ \bibinfo {author} {\bibfnamefont {A.}~\bibnamefont
  {Krasnok}},\ }\bibfield  {title} {\bibinfo {title} {Tunable phase-change
  metasurfaces},\ }\href@noop {} {\bibfield  {journal} {\bibinfo  {journal}
  {Nature Nanotechnology}\ ,\ \bibinfo {pages} {1}} (\bibinfo {year}
  {2021})}\BibitemShut {NoStop}%
\bibitem [{\citenamefont {Hosseini}\ \emph {et~al.}(2014)\citenamefont
  {Hosseini}, \citenamefont {Wright},\ and\ \citenamefont
  {Bhaskaran}}]{hosseini2014optoelectronic}%
  \BibitemOpen
  \bibfield  {author} {\bibinfo {author} {\bibfnamefont {P.}~\bibnamefont
  {Hosseini}}, \bibinfo {author} {\bibfnamefont {C.~D.}\ \bibnamefont
  {Wright}},\ and\ \bibinfo {author} {\bibfnamefont {H.}~\bibnamefont
  {Bhaskaran}},\ }\bibfield  {title} {\bibinfo {title} {An optoelectronic
  framework enabled by low-dimensional phase-change films},\ }\href@noop {}
  {\bibfield  {journal} {\bibinfo  {journal} {Nature}\ }\textbf {\bibinfo
  {volume} {511}},\ \bibinfo {pages} {206} (\bibinfo {year}
  {2014})}\BibitemShut {NoStop}%
\bibitem [{\citenamefont {Schlich}\ \emph {et~al.}(2015)\citenamefont
  {Schlich}, \citenamefont {Zalden}, \citenamefont {Lindenberg},\ and\
  \citenamefont {Spolenak}}]{schlich2015color}%
  \BibitemOpen
  \bibfield  {author} {\bibinfo {author} {\bibfnamefont {F.~F.}\ \bibnamefont
  {Schlich}}, \bibinfo {author} {\bibfnamefont {P.}~\bibnamefont {Zalden}},
  \bibinfo {author} {\bibfnamefont {A.~M.}\ \bibnamefont {Lindenberg}},\ and\
  \bibinfo {author} {\bibfnamefont {R.}~\bibnamefont {Spolenak}},\ }\bibfield
  {title} {\bibinfo {title} {Color switching with enhanced optical contrast in
  ultrathin phase-change materials and semiconductors induced by femtosecond
  laser pulses},\ }\href@noop {} {\bibfield  {journal} {\bibinfo  {journal}
  {ACS photonics}\ }\textbf {\bibinfo {volume} {2}},\ \bibinfo {pages} {178}
  (\bibinfo {year} {2015})}\BibitemShut {NoStop}%
\bibitem [{\citenamefont {Tao}\ \emph {et~al.}(2020)\citenamefont {Tao},
  \citenamefont {Li}, \citenamefont {Wang}, \citenamefont {Wang}, \citenamefont
  {Cai}, \citenamefont {Li}, \citenamefont {Xu}, \citenamefont {Zhang},\ and\
  \citenamefont {Hu}}]{tao2020phase}%
  \BibitemOpen
  \bibfield  {author} {\bibinfo {author} {\bibfnamefont {S.}~\bibnamefont
  {Tao}}, \bibinfo {author} {\bibfnamefont {Q.}~\bibnamefont {Li}}, \bibinfo
  {author} {\bibfnamefont {J.}~\bibnamefont {Wang}}, \bibinfo {author}
  {\bibfnamefont {X.}~\bibnamefont {Wang}}, \bibinfo {author} {\bibfnamefont
  {J.}~\bibnamefont {Cai}}, \bibinfo {author} {\bibfnamefont {S.}~\bibnamefont
  {Li}}, \bibinfo {author} {\bibfnamefont {W.}~\bibnamefont {Xu}}, \bibinfo
  {author} {\bibfnamefont {K.}~\bibnamefont {Zhang}},\ and\ \bibinfo {author}
  {\bibfnamefont {C.}~\bibnamefont {Hu}},\ }\bibfield  {title} {\bibinfo
  {title} {Phase change materials for nonvolatile, solid-state reflective
  displays: From new structural design rules to enhanced color-changing
  performance},\ }\href@noop {} {\bibfield  {journal} {\bibinfo  {journal}
  {Advanced Optical Materials}\ }\textbf {\bibinfo {volume} {8}},\ \bibinfo
  {pages} {2000062} (\bibinfo {year} {2020})}\BibitemShut {NoStop}%
\bibitem [{\citenamefont {Yoo}\ \emph {et~al.}(2016)\citenamefont {Yoo},
  \citenamefont {Gwon}, \citenamefont {Eom}, \citenamefont {Kim},\ and\
  \citenamefont {Hwang}}]{yoo2016multicolor}%
  \BibitemOpen
  \bibfield  {author} {\bibinfo {author} {\bibfnamefont {S.}~\bibnamefont
  {Yoo}}, \bibinfo {author} {\bibfnamefont {T.}~\bibnamefont {Gwon}}, \bibinfo
  {author} {\bibfnamefont {T.}~\bibnamefont {Eom}}, \bibinfo {author}
  {\bibfnamefont {S.}~\bibnamefont {Kim}},\ and\ \bibinfo {author}
  {\bibfnamefont {C.~S.}\ \bibnamefont {Hwang}},\ }\bibfield  {title} {\bibinfo
  {title} {Multicolor changeable optical coating by adopting multiple layers of
  ultrathin phase change material film},\ }\href@noop {} {\bibfield  {journal}
  {\bibinfo  {journal} {Acs Photonics}\ }\textbf {\bibinfo {volume} {3}},\
  \bibinfo {pages} {1265} (\bibinfo {year} {2016})}\BibitemShut {NoStop}%
\bibitem [{\citenamefont {Jafari}\ \emph {et~al.}(2019)\citenamefont {Jafari},
  \citenamefont {Guo},\ and\ \citenamefont
  {Rais-Zadeh}}]{jafari2019reconfigurable}%
  \BibitemOpen
  \bibfield  {author} {\bibinfo {author} {\bibfnamefont {M.}~\bibnamefont
  {Jafari}}, \bibinfo {author} {\bibfnamefont {L.~J.}\ \bibnamefont {Guo}},\
  and\ \bibinfo {author} {\bibfnamefont {M.}~\bibnamefont {Rais-Zadeh}},\
  }\bibfield  {title} {\bibinfo {title} {A reconfigurable color reflector by
  selective phase change of gete in a multilayer structure},\ }\href@noop {}
  {\bibfield  {journal} {\bibinfo  {journal} {Advanced Optical Materials}\
  }\textbf {\bibinfo {volume} {7}},\ \bibinfo {pages} {1801214} (\bibinfo
  {year} {2019})}\BibitemShut {NoStop}%
\bibitem [{\citenamefont {Carrillo}\ \emph {et~al.}(2019)\citenamefont
  {Carrillo}, \citenamefont {Trimby}, \citenamefont {Au}, \citenamefont
  {Nagareddy}, \citenamefont {Rodriguez-Hernandez}, \citenamefont {Hosseini},
  \citenamefont {R{\'\i}os}, \citenamefont {Bhaskaran},\ and\ \citenamefont
  {Wright}}]{carrillo2019nonvolatile}%
  \BibitemOpen
  \bibfield  {author} {\bibinfo {author} {\bibfnamefont {S.~G.-C.}\
  \bibnamefont {Carrillo}}, \bibinfo {author} {\bibfnamefont {L.}~\bibnamefont
  {Trimby}}, \bibinfo {author} {\bibfnamefont {Y.-Y.}\ \bibnamefont {Au}},
  \bibinfo {author} {\bibfnamefont {V.~K.}\ \bibnamefont {Nagareddy}}, \bibinfo
  {author} {\bibfnamefont {G.}~\bibnamefont {Rodriguez-Hernandez}}, \bibinfo
  {author} {\bibfnamefont {P.}~\bibnamefont {Hosseini}}, \bibinfo {author}
  {\bibfnamefont {C.}~\bibnamefont {R{\'\i}os}}, \bibinfo {author}
  {\bibfnamefont {H.}~\bibnamefont {Bhaskaran}},\ and\ \bibinfo {author}
  {\bibfnamefont {C.~D.}\ \bibnamefont {Wright}},\ }\bibfield  {title}
  {\bibinfo {title} {A nonvolatile phase-change metamaterial color display},\
  }\href@noop {} {\bibfield  {journal} {\bibinfo  {journal} {Advanced Optical
  Materials}\ }\textbf {\bibinfo {volume} {7}},\ \bibinfo {pages} {1801782}
  (\bibinfo {year} {2019})}\BibitemShut {NoStop}%
\bibitem [{\citenamefont {de~Galarreta}\ \emph {et~al.}(2020)\citenamefont
  {de~Galarreta}, \citenamefont {Sinev}, \citenamefont {Alexeev}, \citenamefont
  {Trofimov}, \citenamefont {Ladutenko}, \citenamefont {Carrillo},
  \citenamefont {Gemo}, \citenamefont {Baldycheva}, \citenamefont
  {Bertolotti},\ and\ \citenamefont {Wright}}]{de2020reconfigurable}%
  \BibitemOpen
  \bibfield  {author} {\bibinfo {author} {\bibfnamefont {C.~R.}\ \bibnamefont
  {de~Galarreta}}, \bibinfo {author} {\bibfnamefont {I.}~\bibnamefont {Sinev}},
  \bibinfo {author} {\bibfnamefont {A.~M.}\ \bibnamefont {Alexeev}}, \bibinfo
  {author} {\bibfnamefont {P.}~\bibnamefont {Trofimov}}, \bibinfo {author}
  {\bibfnamefont {K.}~\bibnamefont {Ladutenko}}, \bibinfo {author}
  {\bibfnamefont {S.~G.-C.}\ \bibnamefont {Carrillo}}, \bibinfo {author}
  {\bibfnamefont {E.}~\bibnamefont {Gemo}}, \bibinfo {author} {\bibfnamefont
  {A.}~\bibnamefont {Baldycheva}}, \bibinfo {author} {\bibfnamefont
  {J.}~\bibnamefont {Bertolotti}},\ and\ \bibinfo {author} {\bibfnamefont
  {C.~D.}\ \bibnamefont {Wright}},\ }\bibfield  {title} {\bibinfo {title}
  {Reconfigurable multilevel control of hybrid all-dielectric phase-change
  metasurfaces},\ }\href@noop {} {\bibfield  {journal} {\bibinfo  {journal}
  {Optica}\ }\textbf {\bibinfo {volume} {7}},\ \bibinfo {pages} {476} (\bibinfo
  {year} {2020})}\BibitemShut {NoStop}%
\bibitem [{\citenamefont {Hemmatyar}\ \emph
  {et~al.}(2020{\natexlab{e}})\citenamefont {Hemmatyar}, \citenamefont
  {Brown},\ and\ \citenamefont {Adibi}}]{hemmatyar2020tunable}%
  \BibitemOpen
  \bibfield  {author} {\bibinfo {author} {\bibfnamefont {O.}~\bibnamefont
  {Hemmatyar}}, \bibinfo {author} {\bibfnamefont {T.}~\bibnamefont {Brown}},\
  and\ \bibinfo {author} {\bibfnamefont {A.}~\bibnamefont {Adibi}},\ }\bibfield
   {title} {\bibinfo {title} {Tunable ultrahigh-saturation structural colors
  from toroidal resonances by phase-change material sb2s3 metasurfaces},\ }in\
  \href@noop {} {\emph {\bibinfo {booktitle} {CLEO: Applications and
  Technology}}}\ (\bibinfo {organization} {Optical Society of America},\
  \bibinfo {year} {2020})\ pp.\ \bibinfo {pages} {JW2D--33}\BibitemShut
  {NoStop}%
\bibitem [{\citenamefont {Dong}\ \emph {et~al.}(2019)\citenamefont {Dong},
  \citenamefont {Liu}, \citenamefont {Behera}, \citenamefont {Lu},
  \citenamefont {Ng}, \citenamefont {Sreekanth}, \citenamefont {Zhou},
  \citenamefont {Yang},\ and\ \citenamefont {Simpson}}]{dong2019wide}%
  \BibitemOpen
  \bibfield  {author} {\bibinfo {author} {\bibfnamefont {W.}~\bibnamefont
  {Dong}}, \bibinfo {author} {\bibfnamefont {H.}~\bibnamefont {Liu}}, \bibinfo
  {author} {\bibfnamefont {J.~K.}\ \bibnamefont {Behera}}, \bibinfo {author}
  {\bibfnamefont {L.}~\bibnamefont {Lu}}, \bibinfo {author} {\bibfnamefont
  {R.~J.}\ \bibnamefont {Ng}}, \bibinfo {author} {\bibfnamefont {K.~V.}\
  \bibnamefont {Sreekanth}}, \bibinfo {author} {\bibfnamefont {X.}~\bibnamefont
  {Zhou}}, \bibinfo {author} {\bibfnamefont {J.~K.}\ \bibnamefont {Yang}},\
  and\ \bibinfo {author} {\bibfnamefont {R.~E.}\ \bibnamefont {Simpson}},\
  }\bibfield  {title} {\bibinfo {title} {Wide bandgap phase change material
  tuned visible photonics},\ }\href@noop {} {\bibfield  {journal} {\bibinfo
  {journal} {Advanced Functional Materials}\ }\textbf {\bibinfo {volume}
  {29}},\ \bibinfo {pages} {1806181} (\bibinfo {year} {2019})}\BibitemShut
  {NoStop}%
\bibitem [{\citenamefont {Chen}\ \emph {et~al.}(2015)\citenamefont {Chen},
  \citenamefont {Li}, \citenamefont {Zhou}, \citenamefont {Chen}, \citenamefont
  {Luo}, \citenamefont {Liu}, \citenamefont {Zeng}, \citenamefont {Yang},
  \citenamefont {Zhang}, \citenamefont {Han} \emph {et~al.}}]{chen2015optical}%
  \BibitemOpen
  \bibfield  {author} {\bibinfo {author} {\bibfnamefont {C.}~\bibnamefont
  {Chen}}, \bibinfo {author} {\bibfnamefont {W.}~\bibnamefont {Li}}, \bibinfo
  {author} {\bibfnamefont {Y.}~\bibnamefont {Zhou}}, \bibinfo {author}
  {\bibfnamefont {C.}~\bibnamefont {Chen}}, \bibinfo {author} {\bibfnamefont
  {M.}~\bibnamefont {Luo}}, \bibinfo {author} {\bibfnamefont {X.}~\bibnamefont
  {Liu}}, \bibinfo {author} {\bibfnamefont {K.}~\bibnamefont {Zeng}}, \bibinfo
  {author} {\bibfnamefont {B.}~\bibnamefont {Yang}}, \bibinfo {author}
  {\bibfnamefont {C.}~\bibnamefont {Zhang}}, \bibinfo {author} {\bibfnamefont
  {J.}~\bibnamefont {Han}}, \emph {et~al.},\ }\bibfield  {title} {\bibinfo
  {title} {Optical properties of amorphous and polycrystalline sb2se3 thin
  films prepared by thermal evaporation},\ }\href@noop {} {\bibfield  {journal}
  {\bibinfo  {journal} {Applied Physics Letters}\ }\textbf {\bibinfo {volume}
  {107}},\ \bibinfo {pages} {043905} (\bibinfo {year} {2015})}\BibitemShut
  {NoStop}%
\bibitem [{\citenamefont {Ghazi~Sarwat}\ \emph {et~al.}(2019)\citenamefont
  {Ghazi~Sarwat}, \citenamefont {Cheng}, \citenamefont {Youngblood},
  \citenamefont {Sharizal~Alias}, \citenamefont {Sinha}, \citenamefont
  {Warner},\ and\ \citenamefont {Bhaskaran}}]{ghazi2019strong}%
  \BibitemOpen
  \bibfield  {author} {\bibinfo {author} {\bibfnamefont {S.}~\bibnamefont
  {Ghazi~Sarwat}}, \bibinfo {author} {\bibfnamefont {Z.}~\bibnamefont {Cheng}},
  \bibinfo {author} {\bibfnamefont {N.}~\bibnamefont {Youngblood}}, \bibinfo
  {author} {\bibfnamefont {M.}~\bibnamefont {Sharizal~Alias}}, \bibinfo
  {author} {\bibfnamefont {S.}~\bibnamefont {Sinha}}, \bibinfo {author}
  {\bibfnamefont {J.}~\bibnamefont {Warner}},\ and\ \bibinfo {author}
  {\bibfnamefont {H.}~\bibnamefont {Bhaskaran}},\ }\bibfield  {title} {\bibinfo
  {title} {Strong opto-structural coupling in low dimensional gese3 films},\
  }\href@noop {} {\bibfield  {journal} {\bibinfo  {journal} {Nano letters}\
  }\textbf {\bibinfo {volume} {19}},\ \bibinfo {pages} {7377} (\bibinfo {year}
  {2019})}\BibitemShut {NoStop}%
\bibitem [{\citenamefont {Delaney}\ \emph {et~al.}(2020)\citenamefont
  {Delaney}, \citenamefont {Zeimpekis}, \citenamefont {Lawson}, \citenamefont
  {Hewak},\ and\ \citenamefont {Muskens}}]{delaney2020new}%
  \BibitemOpen
  \bibfield  {author} {\bibinfo {author} {\bibfnamefont {M.}~\bibnamefont
  {Delaney}}, \bibinfo {author} {\bibfnamefont {I.}~\bibnamefont {Zeimpekis}},
  \bibinfo {author} {\bibfnamefont {D.}~\bibnamefont {Lawson}}, \bibinfo
  {author} {\bibfnamefont {D.~W.}\ \bibnamefont {Hewak}},\ and\ \bibinfo
  {author} {\bibfnamefont {O.~L.}\ \bibnamefont {Muskens}},\ }\bibfield
  {title} {\bibinfo {title} {A new family of ultralow loss reversible
  phase-change materials for photonic integrated circuits: Sb2s3 and sb2se3},\
  }\href@noop {} {\bibfield  {journal} {\bibinfo  {journal} {Advanced
  Functional Materials}\ ,\ \bibinfo {pages} {2002447}} (\bibinfo {year}
  {2020})}\BibitemShut {NoStop}%
\bibitem [{\citenamefont {Delaney}\ \emph {et~al.}(2021)\citenamefont
  {Delaney}, \citenamefont {Zeimpekis}, \citenamefont {Du}, \citenamefont
  {Yan}, \citenamefont {Banakar}, \citenamefont {Thomson}, \citenamefont
  {Hewak},\ and\ \citenamefont {Muskens}}]{delaney2021non}%
  \BibitemOpen
  \bibfield  {author} {\bibinfo {author} {\bibfnamefont {M.}~\bibnamefont
  {Delaney}}, \bibinfo {author} {\bibfnamefont {I.}~\bibnamefont {Zeimpekis}},
  \bibinfo {author} {\bibfnamefont {H.}~\bibnamefont {Du}}, \bibinfo {author}
  {\bibfnamefont {X.}~\bibnamefont {Yan}}, \bibinfo {author} {\bibfnamefont
  {M.}~\bibnamefont {Banakar}}, \bibinfo {author} {\bibfnamefont {D.~J.}\
  \bibnamefont {Thomson}}, \bibinfo {author} {\bibfnamefont {D.~W.}\
  \bibnamefont {Hewak}},\ and\ \bibinfo {author} {\bibfnamefont {O.~L.}\
  \bibnamefont {Muskens}},\ }\bibfield  {title} {\bibinfo {title} {Non-volatile
  programmable silicon photonics using an ultralow loss sb $ \_2 $ se $ \_3 $
  phase change material},\ }\href@noop {} {\bibfield  {journal} {\bibinfo
  {journal} {arXiv preprint arXiv:2101.03623}\ } (\bibinfo {year}
  {2021})}\BibitemShut {NoStop}%
\end{thebibliography}%
\end{document}